\documentclass[12pt]{article}
\usepackage{a4wide}
\usepackage{graphicx}
\usepackage{amssymb}
\usepackage{amsmath}
\usepackage{authblk}
\usepackage{cite}
\usepackage{braket}

\newcommand{\be}{\begin{equation}}
\newcommand{\ee}{\end{equation}}
\newcommand{\ba}{\begin{eqnarray}}
\newcommand{\ea}{\end{eqnarray}}

\date{}
\title{Scalar Split WIMPs in the Future Direct Detection Experiments}

\author{Karim Ghorbani \thanks{kghorbani@ipm.ir}}   
\affil {\small\it{Physics Department, Faculty of Sciences, Arak University,  \it Arak 38156-8-8349, Iran}}

\author{Hossein Ghorbani\thanks{pghorbani@ipm.ir}}
\affil {\small\it{Institute for Research in Fundamental Sciences (IPM)\\
 \it School of Particles and Accelerators,  \it P.O. Box 19395-5531, Tehran, Iran}}

\begin{document}
 
\maketitle

\abstract{We consider a simple renormalizable dark matter model consisting of two real scalars 
with a mass splitting $\delta$, interacting with the SM particles through the Higgs portal. 
We find a viable parameter space respecting all the bounds
imposed by invisible Higgs decay experiments at the LHC, 
the direct detection experiments by XENON100 and LUX and the 
dark matter relic abundance provided by WMAP and Planck. 
Despite the singlet scalar dark matter model that is fragile against the future direct 
detection experiments, the scalar split model introduced here survives such forthcoming bounds.
We emphasize on the role of the co-annihilation processes and the mixing effects in this feature.   
For $m_{\text{DM}} \sim 63$ GeV in this model we can explain as well the observed gamma-ray excess 
in the analyses of the Fermi-LAT data at Galactic latitudes 
$2^{\circ} \leq |b| \leq 20^{\circ}$ and Galactic longitudes  $|l| < 20^{\circ}$.}

\newpage

\section{Introduction}
\label{sect:introduction}
Although there is no doubt on the existence of dark matter (DM) which is 
forming about $26$ percent 
of the matter content of the Universe \cite{Ade:2013zuv,Hinshaw:2012aka} 
(see e.g. reviews \cite{Bertone:review,Bergstrom:review}), its fundamental interaction 
with ordinary matter of the Standard Model (SM) of particle physics is a tremendous 
mystery in physics today. There is however, a natural explanation for the present value of 
DM relic density in terms of the thermal freeze-out mechanism of weakly interacting massive 
particles (WIMPs). Exploiting the WIMP paradigm, a large number of theories 
beyond the SM is developed with a DM candidate
as a WIMP, we name for instance supersymmetric models with R-parity 
, models with universal extra dimensions, as well as 
models with minimal extension of the SM which is of our interest in this article \cite{Burgess2000,McDonald:1993ex,
Cirelli:2005uq,Pospelov:2011-CPfermionic,LopezHonorez:2012-CPfermionic,
Barger:2008jx,Pospelov:2007mp,Buckley2014}.

All these models can receive stringent constraints on the DM annihilation 
cross section from Planck \cite{Ade:2013zuv} and WMAP \cite{Hinshaw:2012aka},
precise measurements of the DM relic density,
and on the WIMP-nucleon scattering cross section from dark matter experiments
such as LUX \cite{Akerib:LUX} and XENON100 \cite{Aprile:XENON100}.
Moreover, in the case of DM production in particle collider experiments, 
there are measurements such as invisible Higgs decay and missing energy-momentum
that can put further restrictions on the model parameter space 
\cite{Beltran:2010ww,Goodman:2010ku,Bai:2010hh,Belanger:invisible}. 

The new bounds by the coming direct detection experiments such as XENON1T which is going to start
data collection already in this autumn, will certainly exclude many of the current WIMP models. 
The popular singlet scalar dark matter model as the most minimal extension of the SM has been investigated elaborately 
from different points of view in the literature (see for instance \cite{McDonald:1993ex,Burgess2000,ArkaniHamed:2002pa,Barger:2007im}). 
Although this model is fairly successful in various aspects, it is quite in danger 
to be excluded for a wide range of DM mass due to the direct detection experiments 
that will put stringent bounds in the near future \cite{Cline:2013gha}. 
If the direct detection experiments are taken seriously and one is still interested in the scalar extension of the SM, 
the next minimal model that comes to mind is the two real scalar extension dubbed here under the name of
{\it scalar split WIMPs}. We show in this paper that scalar split model is as good as the singlet scalar model with drastically 
improved features in the direct detection part. 

On the other hand, in the light of the recent confirmed observation of the Fermi-LAT extended 
gamma ray excess, many investigations have directed towards possible explanation 
of the gamma excess.
Assuming that the galactic gamma excess produced as a result of DM annihilation
in the galactic center, it is then found in a number of models that DM annihilation 
cross section of order $\sim 10^{-26}$ cm$^3$s$^{-1}$ with DM mass in the range 
$30-50$ GeV can explain the excess, see as examples 
 \cite{Boehm:2014hva,Berlin:2014tja,Boehm:2014bia,Ko:2014gha,Abdullah:2014lla,Berlin:2014pya,Cline:2014dwa,
 Wang:2014elb,Cheung:2014lqa,Balazs:2014jla,Huang:2014cla,Ghorbani:2014qpa,Banik:2014eda
 ,Borah:2014ska,Cahill-Rowley:2014ora,Guo:2014gra,Dolan:2014ska,Biswas:2014hoa,
 Modak:2013jya,Cao:2014efa,Bell:2014xta,Detmold:2014qqa,Cheung:2014tha,Ko:2014loa,
Alvares:2012qv,Basak:2014sza,Martin:2014sxa,Hardy:2014dea}, and 
see \cite{Marshall:2011mm,Okada:2013bna,Liu:2014cma,Freytsis:2014sua,Lacroix:2014eea} for scenarios 
with lighter DM. Later it was found in \cite{Agrawal:2014oha,Calore:2014xka,Calore:2014nla} that 
DM mass of $\sim 35-165$ GeV decaying into $b$ quark pair and DM 
mass large enough to decay into $W^{+}W^{-}$, $ZZ$, $hh$, $\bar tt$ 
pairs can be fitted satisfactorily to the Fermi-LAT data.

In this paper we consider a minimal extension of the SM with two 
additional real scalars denoted by $S_{1}$ and $S_{2}$, which are SM gauge 
singlets and interact with the SM particles via a Higgs portal 
respecting the $\mathbb{Z}_{2}$ symmetry under which the new scalars 
are odd and all the SM particles are even. 
This model suggests two scalar WIMPs with a mass splitting $\delta$  
where only the lighter component is stable and the heaver one is an unstable state \cite{Pospelov:2011-CPfermionic}.      
The viable parameter space constrained by the limits from the observed DM relic abundance, direct detection bounds
as well as invisible Higgs decay width is studied in this work. 
We also show that it is possible to find regions in the viable parameter 
space which can explain the galactic gamma ray excess observed by Fermi-LAT.

The rest of the paper has the following structure. In section~\ref{model} the 
scalar split model is introduced and the relevant free parameters are discussed. 
In section~\ref{relic-density} we discuss how to calculate the relic density using
the DM annihilation cross sections.
Section~\ref{invisible-constrain} is devoted to calculations on the Higgs decay 
to two WIMPs and invisible Higgs decay width is provided in terms of the 
mass range of the DM candidate.
Moreover, the viable parameter space constrained by the DM relic density 
observation as well as invisible Higgs decay width are studied.   
Elastic scattering cross section of DM-nucleon is computed as a function 
of DM mass in section~\ref{direct}, taking into account the limits from 
relic density observation and direct detection experiments.
In section~\ref{gamma-ray-emission} we find how it is possible to explain 
the inner galactic gamma ray excess within the constrained model parameters.
We finish in section~\ref{sect:conclusions} with the conclusion.

\section{Scalar Split WIMPs} 
\label{model}
We consider a renormalizable extension to 
the SM with two new real scalar fields denoted by $S_{1}$ and $S_{2}$. These 
new fields may have small mass splitting and transform 
under $\mathbb{Z}_{2}$ symmetry as $S_{i} \to -S_{i}$.  
The full Lagrangian consists of 
\ba\label{model1}
{\cal L} = {\cal L_{\text{SM}}}+{\cal L}_{\text{DM}}+\cal{L}_\text{int} \,.
\ea 
The Lagrangian $\mathcal{L}_{\text{DM}}$ incorporates only the WIMPs particles as
\ba \label{LDM}
{\cal L_{\text{DM}}}   =  \frac{1}{2} (\partial_{\mu} S_{1})^2 +\frac{1}{2} (\partial_{\mu} S_{2})^2
        -\frac{m^{2}_{1}}{2} S_{1}^2 -\frac{m^{2}_{2}}{2} S_{2}^2- \frac{\lambda_3}{4} S_{1}^4 
        - \frac{\lambda_4}{4} S_{2}^4 \,.                       
\ea
We could in principle have the interaction term $\lambda_s S_1^2 S_2^2$ in the lagrangian (\ref{LDM}). We will turn 
to this point later in this section. 

In addition, respecting the $\mathbb{Z}_{2}$ symmetry, WIMPs interaction with the SM 
particles are considered through a Higgs portal such that  
\ba
\mathcal{L}_\text{int}(S_{1},S_{2},H)= \lambda_{1} S_{1}^2 H^{\dagger}H + \lambda_{2} S_{2}^2 H^{\dagger}H 
+ \lambda_{12} S_{1} S_{2} H^{\dagger}H \,.
\ea
The SM-Higgs potential is also given by    
\ba
V_{H}  =  \mu^{2}_{H} H^{\dagger}H + \lambda_{H} (H^{\dagger}H)^2  \,.
\ea 
The Higgs field is a SM $SU(2)_{L}$ scalar doublet which develops a non-zero 
vacuum expectation value (vev) which results in 
the electroweak spontaneous symmetry breaking. We then parameterize $H$ as 
\ba
H = \frac{1}{\sqrt{2}} \left( \begin{array}{c}
                                0  \\
                                v + h
                       \end{array} \right)\,,
\ea
where $v $ = 246 GeV.

We can choose a basis in which $ \langle S_{1} \rangle = \langle S_{2} \rangle  = 0$.   
The minimization conditions of the total potential are 
\ba
\frac{\partial V}{\partial H}\Big\rvert_{\langle H\rangle = v/\sqrt{2}} = \frac{\partial V}{\partial S_{1}}\Big\rvert_{\langle S_{1}\rangle=  0} 
 =\frac{\partial V}{\partial S_{2}}\Big\rvert_{\langle S_{2}\rangle=  0} = 0  \,.  
\ea 
These conditions provide us with some relations between the parameters. We work 
them out and identify the entries of the mass matrix.
From condition $\frac{\partial V}{\partial H}\big\vert_{\langle H \rangle} = 0$ we get the relation 
\ba
\mu_{H}^2 = -\lambda_{H} v^2 \,.
\ea
From the other two minimization conditions we get no more relation.
We also get the following results for the entries of the mass matrix
\ba
m^{2}_{S_1} = \frac{\partial^2 V}{\partial S_{1}^2} = m_{1}^2 + \lambda_1 v^{2}\,, ~~~
m^{2}_{S_2} = \frac{\partial^2 V}{\partial S_{2}^2} = m_{2}^2 + \lambda_2 v^{2}\,,
\ea
and
\ba
\label{crossterm}
m^{2}_{S_1,S_2} = \frac{\partial^2 V}{\partial S_1 \partial S_2} = \frac{1}{2} \lambda_{12} v^2\,.
\ea
  
We then indicate the two fields $H_{1}$ and $H_{2}$ as the mass eigenstates by introducing 
the mass mixing angle $\theta$  
\ba
H_{1} = \sin \theta~S_1 + \cos \theta~ S_2\,, 
\nonumber\\
H_{2} = \cos \theta~ S_1 - \sin \theta~ S_2\,, 
\ea
where,
\ba
\label{mixing}
\tan \theta = \frac{y}{1+\sqrt{1+y^2}} \,, ~~~ \text{with}~~ y= \frac{2m^{2}_{S_1,S_2}}{m^{2}_{S_2}-m^{2}_{S_1}}\,.
\ea
The two neutral scalars $H_{1}$ and $H_{2}$ have the corresponding mass eigenvalues as
\ba
\label{eigenvalues}
m^{2}_{H_{1},H_{2}} = \frac{m^{2}_{S_{1}}+m^{2}_{S_{2}}}{2}\pm \frac{m^{2}_{S_{2}}-m^{2}_{S_{1}}}{2} \sqrt{1+y^2}\,.
\ea
We assume that $m_{H_{1}} > m_{H_{2}}$ and therefore $H_{2}$ is the stable DM candidate.
It is then possible to obtain relations for $m_{1}$ and $m_{2}$ in terms of physical masses and couplings
\ba
m_{1}^2 = m^{2}_{H_{1}} \sin^2 \theta+  m^{2}_{H_{2}} \cos^2 \theta - \lambda_{1} v^2 \,,
\nonumber \\
m_{2}^2 = m^{2}_{H_{1}} \cos^2 \theta+  m^{2}_{H_{2}} \sin^2 \theta - \lambda_{2} v^2 \,.
\ea
Moreover, from  eq.~(\ref{crossterm}), eq.~(\ref{mixing}) and eq.~(\ref{eigenvalues}) 
we can express the coupling $\lambda_{12}$
in terms of the masses $m_{H_{1}}$ and $m_{H_{2}}$ and the mixing angle $\theta$, 
\ba
\label{coup12}
\lambda_{12} = \frac{2}{v^2} (m_{H_1}^2-m_{H_2}^2)  \sin 2\theta  \,.
\ea

We now turn to the point we made after equation (\ref{LDM}). If we rewrite the lagrangian (\ref{LDM}) in the 
basis of mass eigenstates $H_1$ and $H_2$, then it includes a term for interacting DM with its parter as 
$6\sin^2\theta \cos^2\theta (\lambda_3+\lambda_4) H_1^2 H_2^2$. Taking $m_{H_1}>m_{H_2}$ it means that 
the co-annihilation process $H_2 H_2 \to H_1 H_1$ kinematically is not allowed. 
Adding the $\lambda_s S_1^2 S_2^2$ term do not introduce any new interactions in the 
lagrangian after going to the mass eigenstate basis, although it modifies the strengths by the new coupling $\lambda_s$. 
For instance, the term above changes as 
$\left( 6\sin^2 \theta \cos^2 \theta (\lambda_{3}+\lambda_{4}-12\lambda_s)+6 \lambda_{s}\right) H_{1}^2 H_{2}^2$,
which again do not contribute in the relic density computation. 
Therefore, the term $S_1^2 S_2^2$ merely enlarges the dimension of the parameter space by one. 
To stay in the most minimal scenario possible we assume that $\lambda_s=0$ in this paper. 

We therefore can take in the present model seven independent parameters 
as $m_{H_1}$, $m_{H_2}$, $\lambda_{1}$, $\lambda_{2}$, $\lambda_{3}$, $\lambda_{4}$ and $\theta$, 
while the coupling $\lambda_{12}$ is then fixed by the relations in eq.~(\ref{mixing}) and eq.~(\ref{coup12}). 
The vacuum stability of the total potential restricts the model parameters. In this regards, we
find at tree level the bounds
\ba
m_{1}^2 + m_{2}^2 + (\lambda_{1}+\lambda_{2}) v^2 > 0 \,, 
\nonumber \\
m_{1}^2 m_{2}^2 + (m_{1}^2 \lambda_{1} + m_{2}^2 \lambda_{2}) v^2 +  \lambda_{1} \lambda_{2} v^4 > \lambda_{12}^2 v^4 \,.
\ea
In addition, the perturbativity of the model requires the upper
bounds on the couplings, $|\lambda_{i}| < 4\pi$.

\begin{figure}
\begin{center}
\includegraphics[scale=1,angle =0]{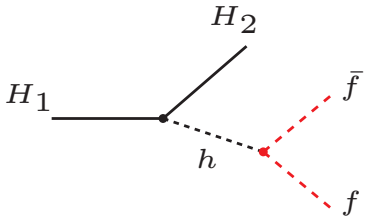}
\end{center}
\caption{Three body decay of the scalar $H_{1}$ into the scalar DM and a fermion pair.}
\label{darkdecay}
\end{figure}

When the small mass splitting is the case then the heavy component WIMP can 
decay into an off-shell Higgs and the light partner as $H_{1} \to H_{2} h$ 
where $h$ itself decays successively into a fermion pair as $h \to \bar f f$. 
The Feynman diagram for the decay is shown in Fig.~\ref{darkdecay}.

It is necessary to have an estimate on the life time of the heavy component 
over the restricted parameter space to know whether or not it has any contribution
on the DM relic abundance. We provide here the formula of the double 
differential partial decay width for
$H_{1}(k) \to H_{2}(p_{3})~ \bar f(p_{1})~ f(p_{2})$ 
\ba
\frac{d^{2}\Gamma}{dt~du} = \frac{3m_{f}^{2} [(\lambda_{1}-\lambda_{2}) \sin 2\theta+
   \lambda_{12} \cos 2\theta~]^2}{128\pi^3m_{H_1}^3} 
 \Big[ \frac{t+m_{h}^{2}-m_{H_{2}}^2-4m_{f}^2}{(t-m_{h}^2)^2+\Gamma_{h}^{2}m_{h}^{2}} \Big]\,,
\label{lifetime}
\ea  
where the mandelstam variables are $t = (p_{1}+p_{2})^2$ and $u = (p_{2}+p_{3})^2$.

\section{Dark Matter Relic Abundance}
\label{relic-density}

\begin{figure}
\begin{center}
\includegraphics[scale=1,angle =0]{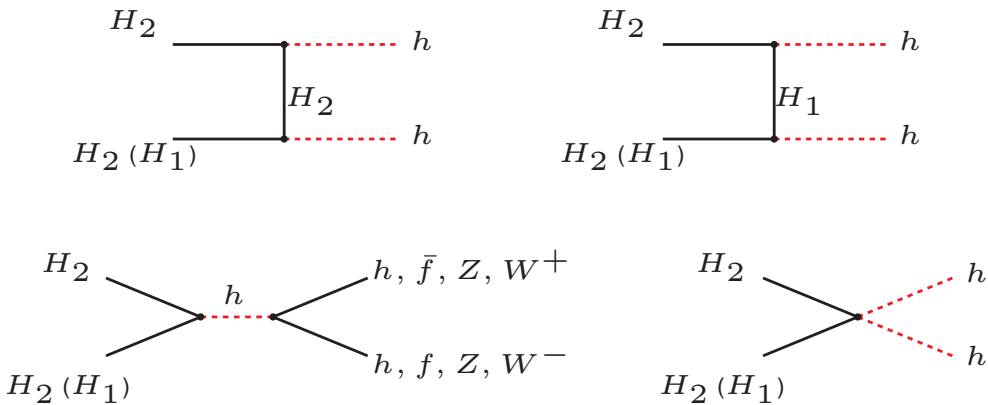}
\end{center}
\caption{The Feynman diagrams for the DM (co)-annihilation into SM final states. 
Diagrams with more than two particles in the final state are not shown.}
\label{darkanni}
\end{figure}

Assuming that DM particles have been in thermal equilibrium in the early Universe, 
the present density of DM depends somehow on the so-called freeze-out 
temperature, $T_{f}$, the epoch in which dark particles become non-relativistic 
and go out of the equilibrium. At freeze-out temperature the annihilation rate of DM 
falls off below the Hubble expansion
rate. On the other side, due to the low budget of the kinetic energy, 
the DM production reactions get suppressed. The relic density of DM is 
computed by solving the Boltzmann equation(s) for the time evolution of DM 
number density, $n_{\text{DM}}$. In the model under consideration, there are 
two new scalars beside the SM particles that their number density evolutions 
are relevant in order to obtain the DM relic abundance.         
We assume that $H_{2}$ is the lighter component and thus is stable.
We therefore consider $H_{2}$ as our DM candidate with mass $m_{H_{2}}$ 
that $m_{H_{1}}> m_{H_{2}}$. So the heavier scalar $H_1$ can undergo 
the decay $H_{1} \to H_{2}+\text{SM}$. 

Annihilation reactions are one type of processes that change the number 
density ($n_{1}$ and $n_{2}$) of our species here. 
The possible annihilations of $H_{1}$ and $H_{2}$ to SM particles are depicted in 
Fig.~\ref{darkanni}. As it is evident from the Feynman diagrams, 
annihilation reactions into SM fermion pairs, $W^{+}W^{-}$ and $ZZ$ occur 
via  s-channel while annihilation 
into SM-Higgs pair is possible through s-, t- and u-channel.   
An annihilation process in which DM particle annihilates together 
with $H_{1}$ is the so-called {\it co-annihilation} reaction. Another type of reaction that 
changes the number density is the decay process of the heavier component, i.e., $H_{1}$. 
In principle, the abundance of $H_{1}$ and $H_{2}$ are determined by solving two 
coupled Boltzmann equations. The two Boltzmann equations can be written in a single Boltzmann 
equation with an effective (co-)annihilation cross section \cite{Griest:1990kh,Edsjo:1997bg,Belanger:2004yn}, 

\begin{equation}
 \frac{dn}{dt}=-3Hn-\braket{\sigma_{\text{eff}}\,v}\left(n^2 - n^2_{eq} \right)\, ,
\end{equation}
where $n \equiv n_{H_1}+n_{H_2}$ and 
\begin{equation}
\sigma_{\text{eff}}=\frac{1}{g_{\text{eff}}}\left(\sigma_{22}+\sigma_{11}\left(1+\frac{\delta}{m_{H_2}}\right)^{3}e^{-2\delta/T}
+2\sigma_{12}\left(1+\frac{\delta}{m_{H_2}}\right)^{3/2}e^{-\delta/T} \right)\, ,
\end{equation}
where $\sigma_{22},\sigma_{11}$ and $\sigma_{12}$ stand for (co-)annihilation processes 
$H_2 H_2 \to \text{SM}\, \text{SM}$,~$ H_1 H_1 \to \text{SM}\, \text{SM}$ and $H_2 H_1 \to \text{SM}\, \text{SM}$ respectively 
with $g_{\text{eff}}=1+\left(1+\frac{\delta}{m_{H_2}}\right)^{3/2}e^{-\delta/T}$.

The expression $\braket{ \sigma_{\text{eff}} \, v} $ indicates thermal average over effective annihilation 
cross section $\times$ relative velocity at temperature $T$.
In appendix A we present the formulas for annihilation cross sections of dark 
matter candidate in four possible channels. To confirm our analytical formula we 
employ the program CalcHEP \cite{Belyaev:CalcHEP} which in turn requires implementation of 
our model into the program LanHEP \cite{Semenov:LanHEP}. 
To perform the analysis for the DM relic abundance we need to 
solve numerically the Boltzmann equation. To this end, we
utilize the program MicrOMEGAs \cite{Belanger:MICRO} for our model.
\begin{figure}
\begin{center}
\includegraphics[scale=.45,angle =-90]{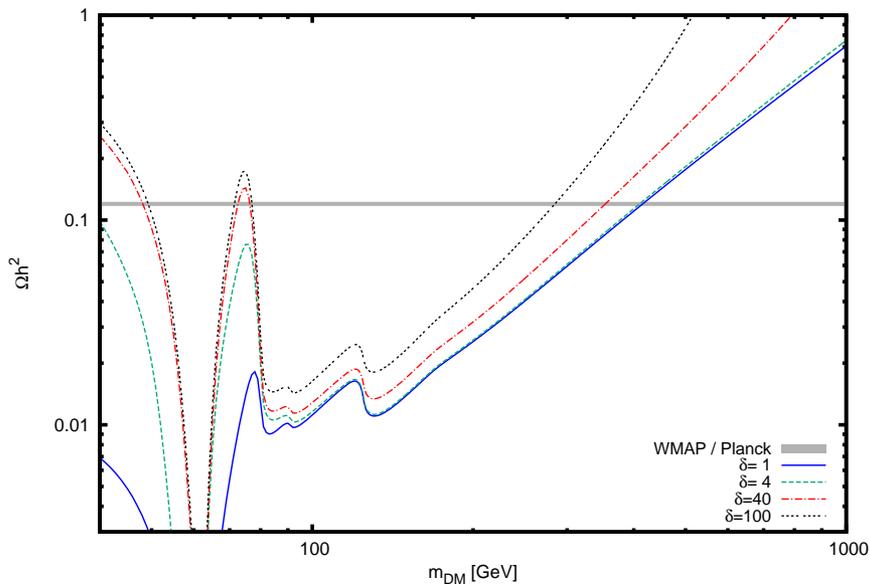}
\end{center}
\caption{The dependency of the relic density on the mass splitting $\delta$ for a 
wide range of DM mass.}
\label{coannihilation}
\end{figure}

As explained earlier we have two choices for a set of 
independent parameters we would like to place the constraints on. 
Notice that the couplings $\lambda_{3}$ and $\lambda_{4}$ do not show up 
in DM annihilation cross sections at the tree level, however these couplings appear through
the strength of the vertex $H_{1}^2 H_{2}^2$. 
We fix the two couplings as $\lambda_{3} = \lambda_{4} = 0$. 
Therefore, one possibility is choosing the set of 
parameters \{$m_{H{1}}$, $m_{H_{2}}$, $\lambda_{1}$, $\lambda_{2}$, $\lambda_{12} $\} and 
the other option is the set \{$m_{H_{1}}$, $m_{H_{2}}$, $\lambda_{1}$, $\lambda_{2}$, $\theta$\}.         
In our analysis we choose the second set and apply the relation in eq.~(\ref{coup12})
to obtain the coupling $\lambda_{12}$ by fixing the mixing angle $\theta$. 

Let us define the mass splitting as $\delta\equiv\Delta m_{12} = m_{H_{1}} - m_{H_{2}}$.
Taking into account the co-annihilation processes, we check numerically 
the dependency of the relic density on the mass splitting $\delta$. 
For a point in the parameter space with $\lambda_{1} = 0.56$, $\lambda_{2} = 0.33$ and $\sin \theta = 0.1$,
the results are compared in Fig.~\ref{coannihilation} 
for $\delta = 1$ GeV, $4$ GeV, $40$ GeV and $100$ GeV. 
Since the co-annihilation effects are larger for smaller value of $\delta$, 
for the present model, it is evident from Fig~.\ref{coannihilation} that 
the relic density is reduced by the co-annihilation effects.

\section{Invisible Higgs Decay}
\label{invisible-constrain}
The DM candidate in the scalar split model interacts with the SM particles via SM-Higgs mediator. It also opens up 
the possibility for the 125 GeV Higgs to decay into the new scalars. Constraints on the model parameters 
are placed by requiring the invisible Higgs decay to be consistent with the Large Hadron Collider 
(LHC) measurements.  
The total decay width of 125 GeV Higgs decaying into SM particles is $\sim$ 4.1 MeV \cite{Heinemeyer:2013} 
which get enhanced by three invisible decay width of the SM-Higgs, $h \to H_{1} H_{1}$,  
$h \to H_{1} H_{2}$ and  $h \to H_{2} H_{2}$. Given an experimental upper limit for the invisible 
branching ratio for the Higgs boson as $\Gamma_{\text{inv}}/(\Gamma_{\text{inv}}+\Gamma_{\text{v}})\sim$ 0.35 
\cite{Belanger:invisible} we 
put a bound on the total invisible decay 
width as $\Gamma^{\text{total}}_{\text{inv}} <$ 2.15 MeV. On the other hand, the total invisible decay width 
in this model is saturated by three possible decays of the Higgs:  
\ba
\Gamma^{11}_{\text{inv}} (h \to H_{1} H_{1}) =
\frac{v^2 (\lambda_{1}\sin^2 \theta+\lambda_{2}\cos^2 \theta+\lambda_{12} \sin \theta \cos \theta)^2}{8\pi m_{h}} 
(1-\frac{4 m_{H_{1}}^{2}}{m_{h}^{2}})^{1/2} \,,
\ea
\ba
\Gamma^{22}_{\text{inv}} (h \to H_{2} H_{2}) =
\frac{v^2 (\lambda_{1}\cos^2 \theta+\lambda_{2}\sin^2 \theta - \lambda_{12} \sin \theta \cos \theta)^2}{8\pi m_{h}} 
(1-\frac{4 m_{H_{2}}^{2}}{m_{h}^{2}})^{1/2} \,,
\ea
and
\ba
&&\hspace{-3.3cm} 
\Gamma^{12}_{\text{inv}} (h \to H_{1} H_{2}) =
\frac{v^2 [(\lambda_{1}-\lambda_{2}) \sin 2 \theta + \lambda_{12} \cos 2 \theta]^2}{8\pi m_{h}^{3}} \times
\nonumber\\&&\hspace{-.8cm}
[m_{h}^{2}-(m_{H_{1}}^{2}+m_{H_{2}}^{2})^2]^{1/2}[m_{h}^{2}-(m_{H_{1}}^{2}-m_{H_{2}}^{2})^2]^{1/2}  \,.
\ea
The invisible Higgs decay width depends 
on DM mass $m_{H_{2}}$ and $\delta$ 
as the following: 
\ba
\Gamma^{\text{total}}_{\text{inv}}  &&\hspace{-0.5cm}= \Gamma^{22}_{\text{inv}} \hspace{3cm}  \text{when}   \hspace{1cm}  \frac{m_{h}}{2} - \frac{\delta}{2} < m_{H_{2}} < \frac{m_{h}}{2}\,,
\nonumber \\&&\hspace{-1.55cm}
\Gamma^{\text{total}}_{\text{inv}}  = \Gamma^{22}_{\text{inv}}+\Gamma^{12}_{\text{inv}} \hspace{2cm}  \text{when}  \hspace{1cm}  \frac{m_{h}}{2}-\delta <  m_{H_{2}} < \frac{m_{h}}{2} - \frac{\delta}{2} \,,
\nonumber \\&&\hspace{-1.55cm}
\Gamma^{\text{total}}_{\text{inv}}  = \Gamma^{22}_{\text{inv}}+\Gamma^{12}_{\text{inv}}+\Gamma^{11}_{\text{inv}} \hspace{.7cm}  \text{when}   \hspace{2.5cm}    m_{H_{2}} < \frac{m_{h}}{2} - \delta \,.
\ea

\begin{figure}
\begin{minipage}{0.35\textwidth}
\includegraphics[width=\textwidth,angle =-90]{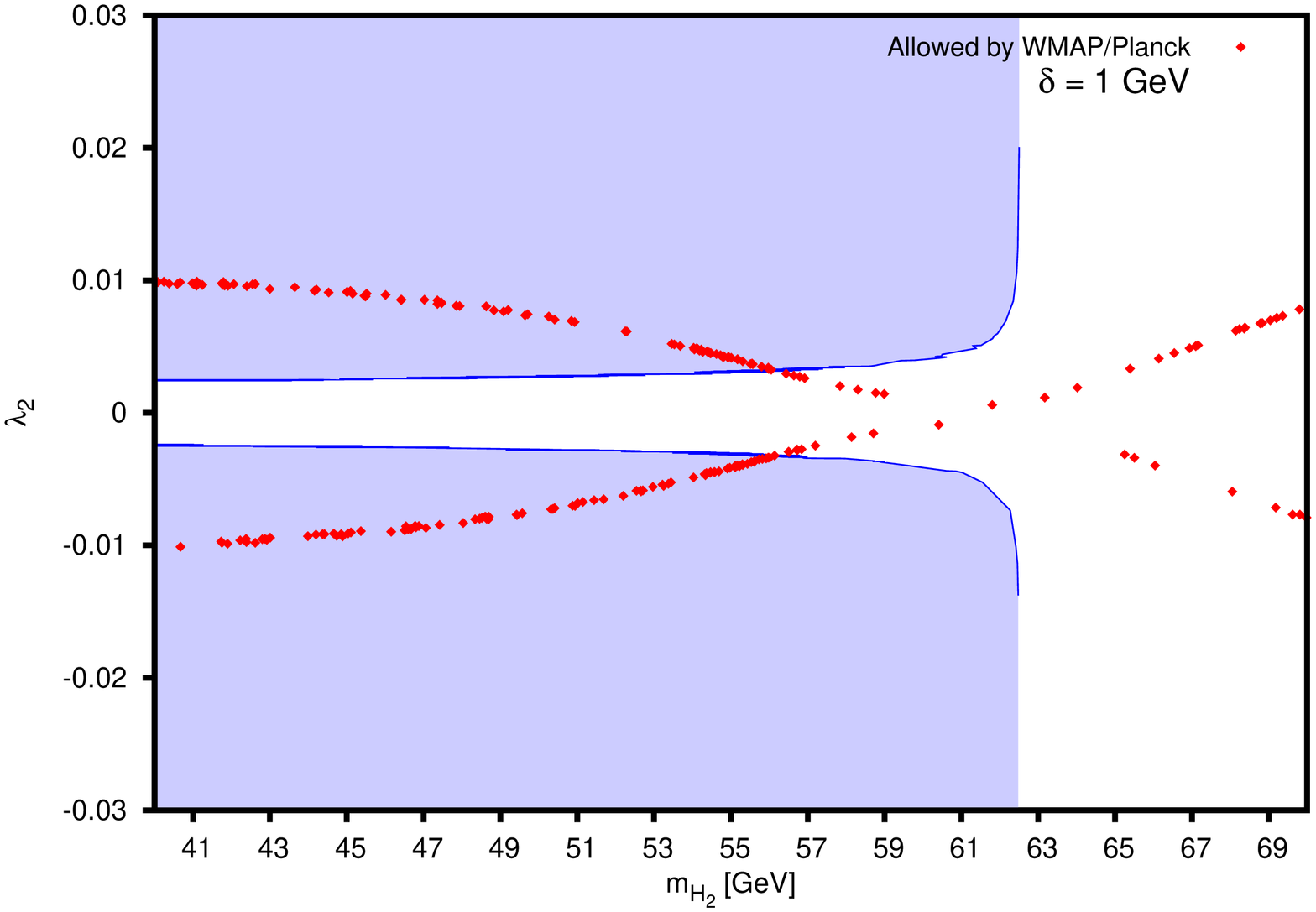}
\end{minipage}
\hspace{2cm}
\begin{minipage}{0.35\textwidth}
\includegraphics[width=\textwidth,angle =-90]{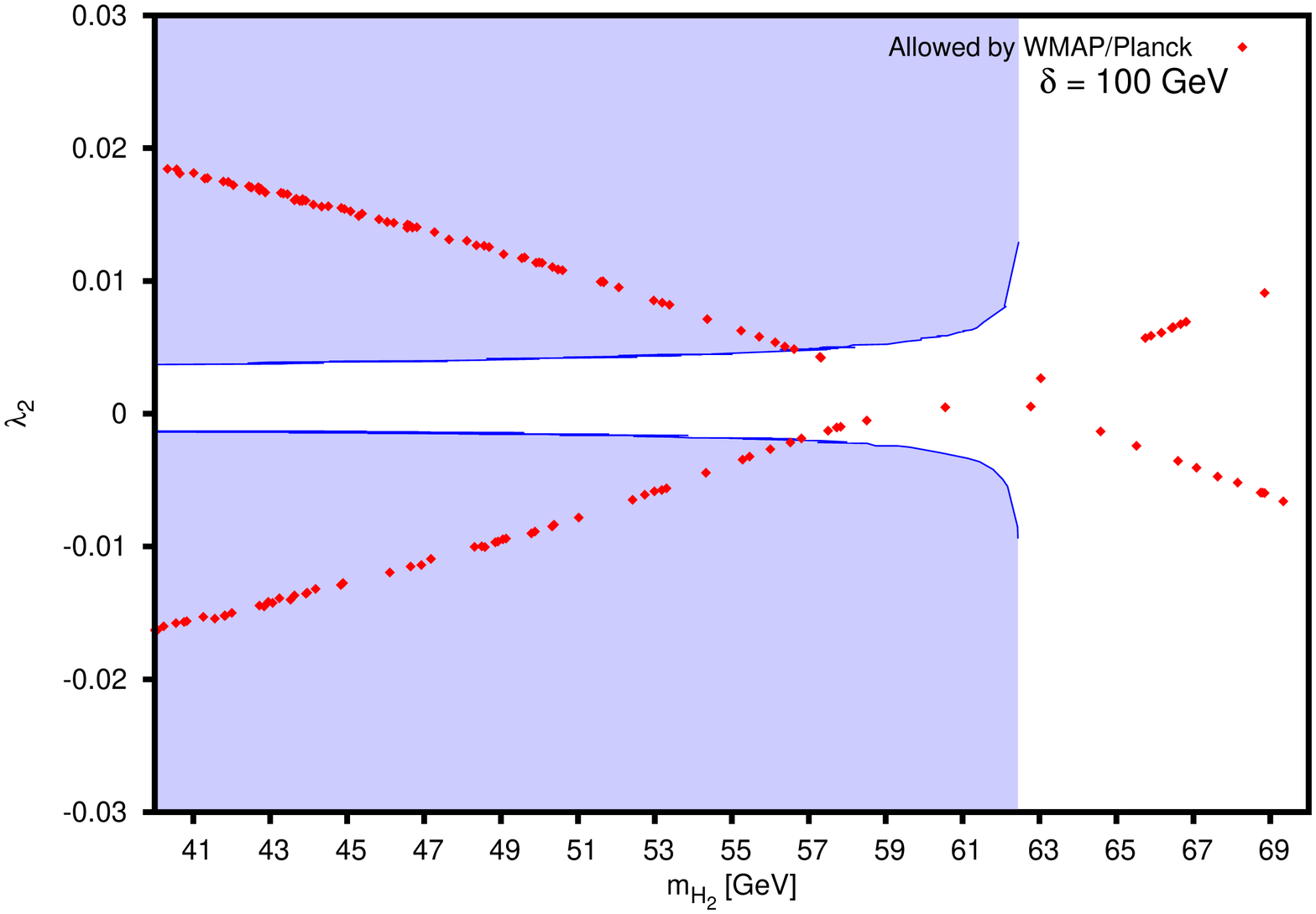}
\end{minipage}
\caption{Shown are the allowed DM mass in the viable parameter space respecting
the relic density and invisible Higgs decay width
constraints for {\it left}) $\delta=1$ GeV {\it right}) $\delta=100$ GeV.}
\label{invisrelic}
\end{figure}
  
Let us now begin with our probe over the parameter space of the model. 
To proceed we put together the constraints imposed on the parameter 
space from relic density analysis and the invisible Higgs decay width. 
This has been done for two values of the mass splitting $\delta=1$ GeV and $\delta=100$ GeV
in order to investigate the role of this parameter on the viable space and 
confronting that with the singlet scalar dark matter case.

We have generated random values for DM mass with 40 GeV $< m_{H_{2}} < m_{h}/2$,
$-1< \lambda_{1} < 1$, and taking $\lambda_2=\lambda_1/5$ and $\sin{\theta}=0.1$.
Using the combined results from WMAP and Planck for the present DM relic density, 
the results exhibited in Fig.~\ref{invisrelic} show the 
viable parameter space for different values of the  
mass splitting parameter $\delta$. The region colored in blue is excluded by the invisible Higgs decay width. 
There are some comments in order for the Fig.~\ref{invisrelic}:

First, for DM mass below $m_{h}/2$, only DM
annihilation into fermions mediated via SM-Higgs are potentially allowed processes, thus 
one expects enhancement on the cross section near the SM-Higgs mass resonance.  
It is evident from Fig. \ref{invisrelic} that the resonance occurs around DM mass $m_{H_2}\sim m_h/2 \sim 62$ GeV 
where the coupling $\lambda_2$ (as well as the annihilation cross section) takes its smallest value,
while $\lambda_2$ grows up for DM masses smaller or greater than the resonance mass $62$ GeV.
Moreover, it is seen that for both mass splittings $\delta =1$ GeV and $\delta=100$ GeV all DM masses smaller 
than $m_{H_2}\sim 56$ GeV are excluded. Finally, the range of the allowed DM mass in the scalar split 
model is almost the same as that of the single scalar dark matter model 
where $m_{\text{DM}} < 55 $ GeV is excluded by the LHC bounds on the invisible Higgs decay width \cite{Cline:2013gha}.

\section{Direct Detection}
\label{direct}

Direct detection experiments are designed to study the unknown nature of DM 
interaction with ordinary matter. In these experiments the attempt is 
to measure the enticing event rate for the DM scattering off the target nuclei 
in the detector. Although the present results from DM experiments such as 
LUX \cite{Akerib:LUX} and XENON100 \cite{Aprile:XENON100} show no evidence for DM interactions, 
they offer an impressive upper bound on the spin-independent 
DM-nucleon elastic scattering cross section. We will apply these findings 
in the following to constrain further the parameter 
space of our model which is already restricted by the limits from WMAP and Planck. 
\begin{figure} 
    \centering
     \includegraphics[scale=.8,angle=0]{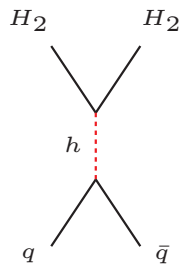}
    \caption{The relevant Feynman diagram for the WIMP-nucleon elastic scattering.}
 \label{direct-diagrams}
 \end{figure}
To this end, we need to calculate the elastic scattering of WIMP-nucleon. In the present
particular model the interaction of DM with nucleon occurs through a 
fundamental interaction of DM with quark which is mediated by the SM-Higgs, where
the relevant Feynman diagram is depicted in Fig.~\ref{direct-diagrams}. 
The effective Lagrangian responsible for the DM-quark interaction is,
\ba
{\cal L}_{\text{eff}} = \alpha_{q} H_{2} H_{2}~\bar q q \,,   
\ea
where, the coupling $\alpha_{q}$ is given by 
\ba\label{alphaq}
\alpha_{q} = \frac{m_{q}}{m_{h}^2} 
(\lambda_{1}\cos^2 \theta+\lambda_{2}\sin^2 \theta - \lambda_{12} \sin \theta \cos \theta).
\ea         
To find the elastic scattering cross section we can invoke the assumption 
that in the limit of vanishing momentum transfer it is possible to replace the
nucleonic matrix element including quark current with that 
containing nucleon current up to some proportionality 
factor \cite{Belanger:2008-Direct,Ellis:2008,Nihei:2004,Ellis:2000}, 
see also \cite{Crivellin:2013}. We arrive at the final result for the 
spin-independent (SI) cross section of DM-nucleon as
\ba\label{sigmaSI}
\sigma^{\text{N}}_{\text{SI}} = 
\frac{\alpha_{N}^2 \mu_{N}^2}{\pi m_{\text{DM}}^2}\,,
\ea
where $\mu_{N}$ is the reduced mass of the DM-nucleon system and the factor 
$\alpha_{N}$ depends on the scalar couplings $f^{N}_{Tq}$ and $f^{N}_{Tg}$ as
\ba
\alpha_{N} = m_{N} \sum_{q = u,d,s} f^{N}_{Tq} \frac{\alpha_{q}}{m_{q}} 
+ \frac{2}{27} f^{N}_{Tg} \sum_{q = c,b,t}   \frac{\alpha_{q}}{m_{q}} \,.
\ea
In our numerical calculations we use the following values for the scalar couplings
\ba
f^{p}_{u} = 0.0153,~~~~~~~ f^{p}_{d} = 0.0191, ~~~~~~~ f^{p}_{s} = 0.0447 \,.
\ea

\begin{figure}
\begin{minipage}{0.39\textwidth}
\includegraphics[width=\textwidth,angle =-90]{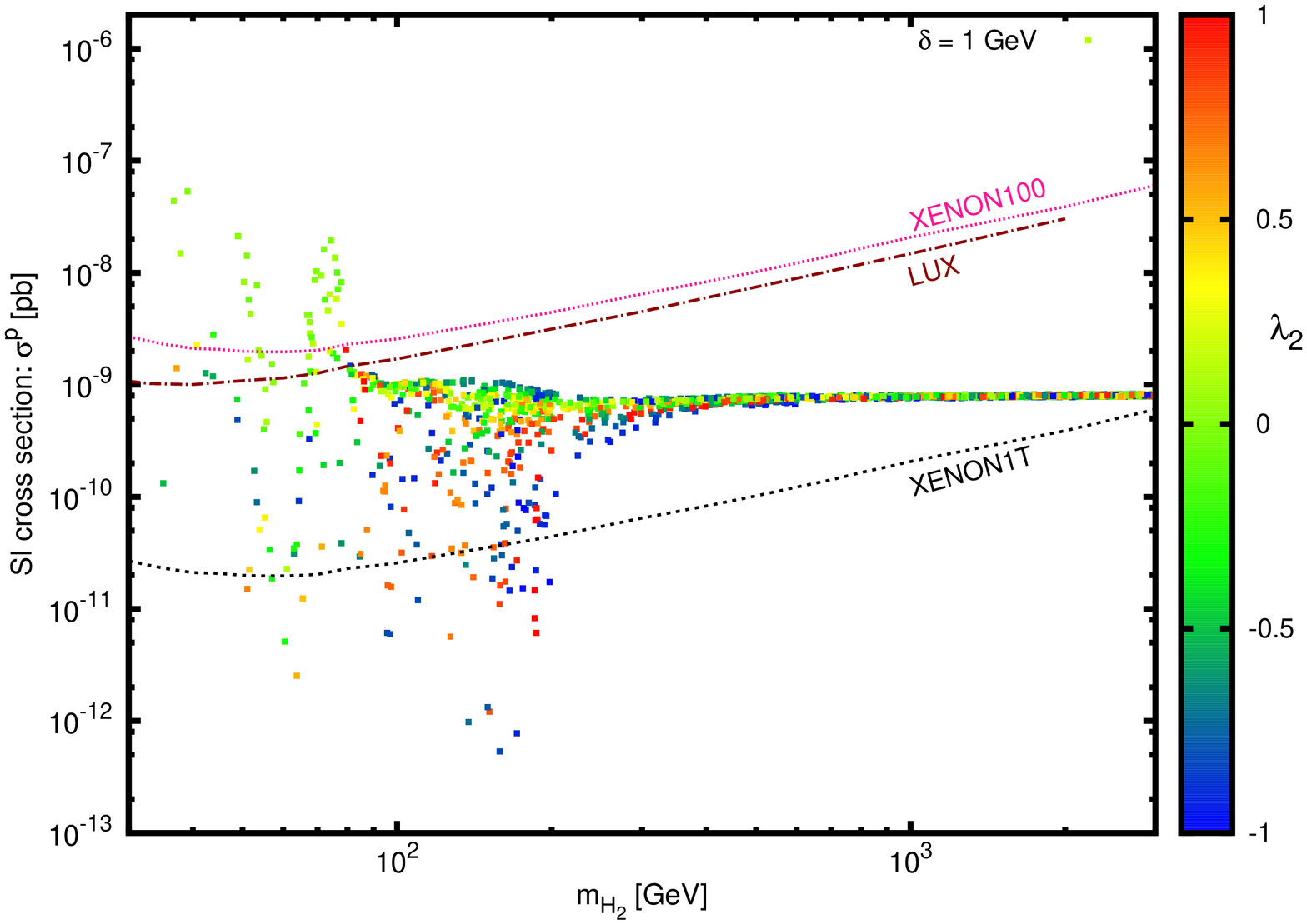}
\end{minipage}
\hspace{2cm}
\begin{minipage}{0.39\textwidth}
\includegraphics[width=\textwidth,angle =-90]{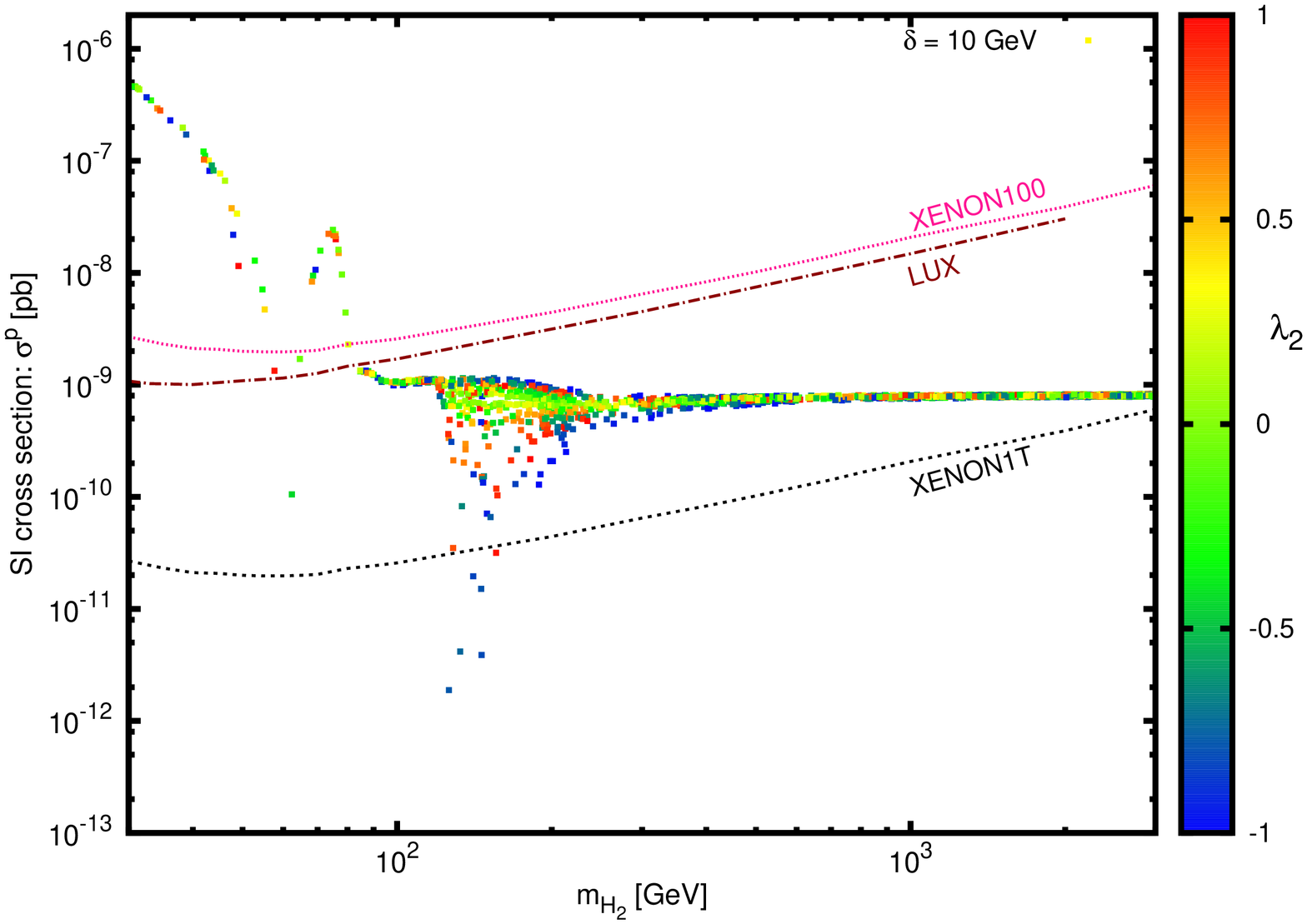}
\end{minipage}
\caption{Spin-independent DM-nucleon elastic scattering cross section
are shown as a function of the DM mass and comparison has made 
with the latest results from LUX and XENON100 experiments and  
the future experiment XENON1T. 
In the left panel the mass splitting is $\delta = 1$ GeV and 
in the right panel $\delta = 10$ GeV. The vertical color spectrum indicates
the size of $\lambda_{2}$. We have chosen for the couplings as $-1 < \lambda_1 < 1$ and $-1 < \lambda_2 < 1$. }
\label{dirdelta1-10}
\end{figure}

\begin{figure}
\begin{minipage}{0.39\textwidth}
\includegraphics[width=\textwidth,angle =-90]{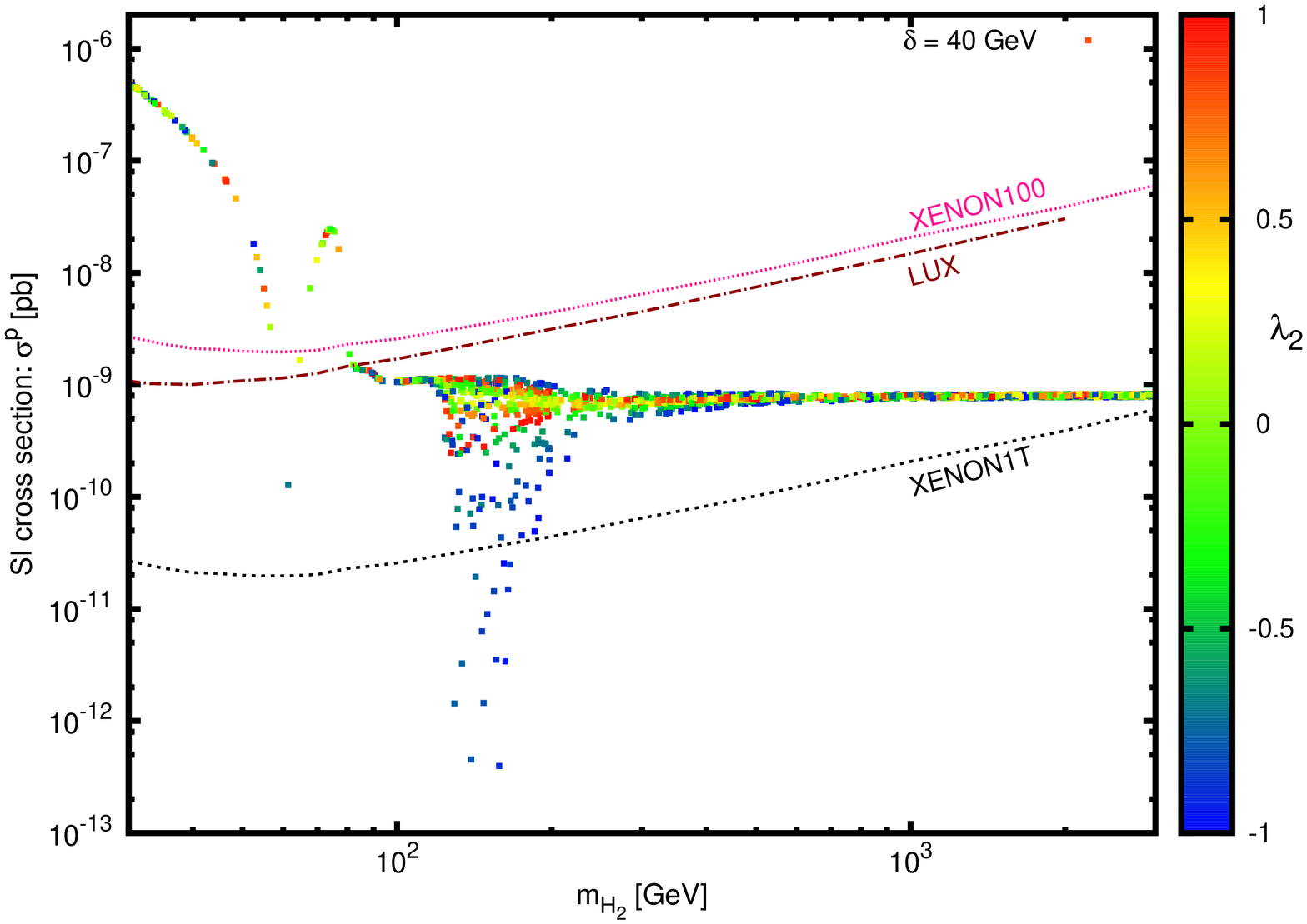}
\end{minipage}
\hspace{2cm}
\begin{minipage}{0.39\textwidth}
\includegraphics[width=\textwidth,angle =-90]{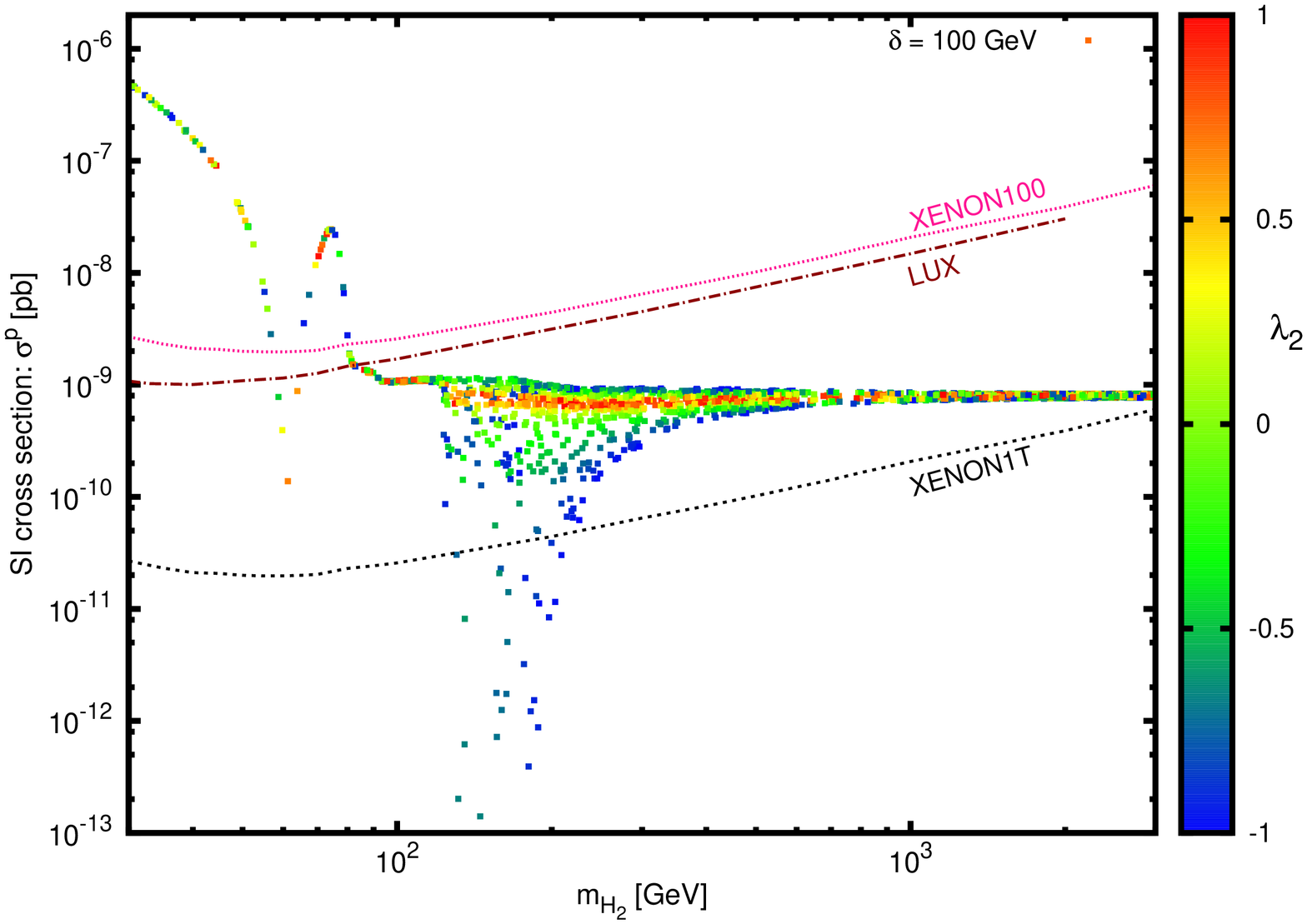}
\end{minipage}
\caption{The same as in Fig.~\ref{dirdelta1-10} with the left panel 
for the mass splitting $\delta = 40$ GeV and the right 
panel for $\delta = 100$ GeV.}
\label{dirdelta40-100}
\end{figure}

We compute the elastic scattering cross section for about $10^6$ points in 
the parameter space by random generation of the relevant parameters 
with $-1 < \lambda_1 < 1 $, $-1 < \lambda_2 < 1 $, 40 GeV $< m_{\text{DM}} < 3$ TeV
and $\sin{\theta} = 0.1$.  
We report on our numerical results for the elastic scattering cross section of DM-proton
in Fig.~\ref{dirdelta1-10} and Fig.~\ref{dirdelta40-100}, considering 
four different mass splittings in the model, namely $\delta = 1, 10, 40, 100$ GeV.  
We have exploited the viable parameter space fulfilling already 
the relic density bound
to obtain numerically the elastic scattering cross sections for a wide range of DM mass. These
results have been plotted against the experimental bounds provided by LUX and XENON100 and 
the estimated bound for the future XENON1T.

In all plots it can be seen easily that the cross section falls off suddenly at the Higgs mass 
resonance region as pointed out before in section \ref{invisible-constrain}. It is for 
the resonance mass, i.e., around $62$ GeV and for $m_{\text{DM}}\gtrsim  4$ TeV that the singlet scalar model \cite{Cline:2013gha} 
can evade the future direct detection bounds. The success for the scalar split model is that 
it not only can evade the LUX and the XENON100 constraints but there are many DM candidates for which  
the values of the elastic scattering cross sections go even much below the future direct detection experiments
such as XENON1T for a quite wide range of DM mass. This feature is true for all mass splittings 
$\delta$. However, the viable parameter space is slightly sensitive to the mass splitting as seen 
in Figs.~ \ref{dirdelta1-10} and \ref{dirdelta40-100}.  When $\delta$ is small, i.e., when
DM scalar and its partner in the model have more or less the same mass, the viable DM mass 
can be in the range $\sim 50-200$ GeV for $\delta = 1$ GeV if we consider the constraints imposed by XENON1T. 
With increasing the mass splitting $\delta$, the viable DM mass is limited to values in 
the range $\sim 125-200$ GeV. It seems therefore that for small enough $\delta$ the 
parameter space is greater than that with much bigger $\delta$. Looking at Fig. 
\ref{coannihilation}, note that for 
the viable space, i.e., for $m_{\text{DM}}\sim 125-200$ GeV, the relic density does not change 
considerably going from $\delta=1$ GeV to $\delta=100$ GeV. On the other hand, the value of 
the splitting mass $\delta$ can make a big change in the relic density for $m_{\text{DM}}\lesssim 60$ GeV
and $m_{\text{DM}}\gtrsim 400$ GeV.

\begin{figure}
\begin{center}
\includegraphics[scale=.45,angle =-90]{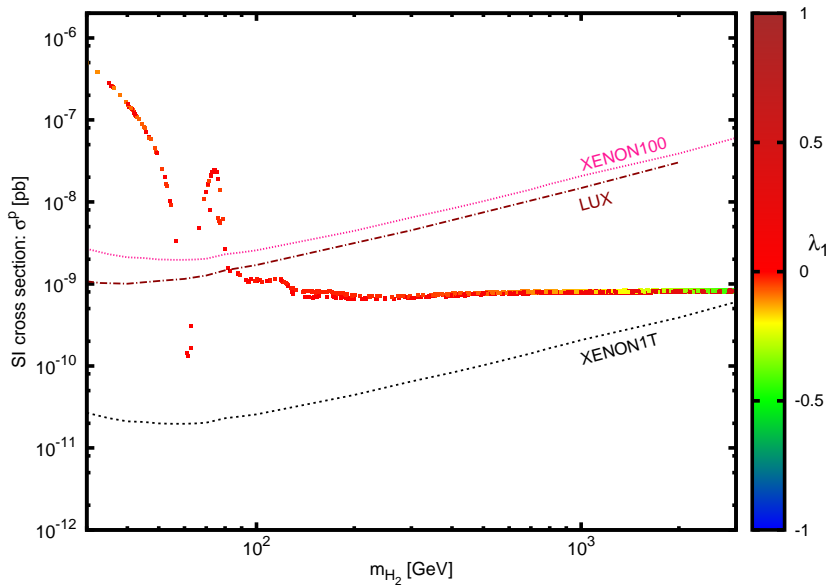}
\end{center}
\caption{The elastic scattering cross section with $\sin\theta=0$ for the parameter space bounded by the 
present Universe DM relic density.}
\label{teta-zero}
\end{figure}

We claim that the improvement in the spin-independent DM-nucleon scattering cross section for the present
scalar split model of DM stems from two distinct effects. One is the contribution of 
the co-annihilation processes and the other one is what we call it {\it mixing effect}. 
These two effects do not exist in the singlet scalar DM model. 
To clarify this statement, we repeat the computation for the 
DM-nucleon scattering cross section for $\sin\theta=0$ where the contribution from the co-annihilation
processes in the relic density and the mixing effect are both absent. 
The results for $\sigma_{\text{SI}}$ that respect the relic
density bound is plotted in Fig.~\ref{teta-zero}.
This figure actually accounts for the similar results presented in \cite{Cline:2013gha} 
for the singlet scalar DM model. It is evident from Fig.~\ref{teta-zero}
that the feature in Fig.~\ref{dirdelta1-10} and \ref{dirdelta40-100} disappears in the case $\sin\theta=0$.
There is a simple explanation why the new feature is not possible in the single scalar model. 
The reason hinges in the fact that in this case, both the annihilation cross section and the DM-nucleon
scattering cross section are proportional to one common parameter, $\lambda_1$, as can be seen from 
the formulas provided in the appendix \ref{app}. So that it is not possible to get simultaneously 
a quite small value for the DM-nucleon scattering cross section and a large enough annihilation cross section
suitable for predicting the observed relic density.    

The question we would like to address here is that why the new feature for large $\delta$ starts appearing in the 
region with $m_{\text{DM}} \gtrsim 125$ GeV when $\theta \ne 0$. 
We know that for $m_{\text{DM}} < 125$ GeV, only the processes, $H_2 H_2 \to \bar f f, W^+ W^-, ZZ$
contribute to the total annihilation cross section. Looking at the relevant formulas 
given in the appendix \ref{app}, we find out that both the annihilation cross section and DM-nucleon
scattering cross section are proportional to one common parameter, $\beta$, where 
$\beta = \lambda_{1}\cos^2 \theta+\lambda_{2}\sin^2 \theta - \lambda_{12} \sin \theta \cos \theta$.
Therefore, we can apply the same line of reasoning as we did in the singlet model to understand
why in the region with $m_{\text{DM}} < 125$ GeV we see the same prediction as 
the one in the singlet model. Of course, for small mass splitting, i.e., $\delta = 1$ GeV
the co-annihilation effects are sizable such that for $m_{\text{DM}} < 125$ GeV, the spilt scalar
model and singlet model show different predictions for the DM-nucleon scattering cross section.
Now, when $m_{\text{DM}}$ gets values larger than $125$ GeV, the process $H_2 H_2 \to h h$ starts 
dominating the total annihilation cross section. Here we expect the mixing effects show up.
In this region it is totally possible to find
quite small values for $\beta$ and hence small DM-nucleon scattering cross section. 
At the same time having large values for $\alpha$ results in large enough annihilation cross 
section, where $\alpha = (\lambda_{1}-\lambda_{2}) \sin 2\theta+\lambda_{12} \cos 2\theta$.
To see this point, we need to look at eq.~(\ref{ann3}) in the appendix \ref{app}. We see that even when $\beta$ 
is small, the annihilation cross section can be large enough since $\alpha$ is not necessarily
small and terms involving $\alpha$ will dominate the annihilation cross section.   
We have justified this latter claim in our numerical computations. 
When DM mass is larger than $\sim 188$ GeV such that the process $H_2 H_2 \to h h h$ 
becomes kinematically possible, the aforementioned mixing effects discussed above are plausible, however,
their strength would depend on the size of the couplings $\lambda_1$, $\lambda_2$ and $\delta$.
Heavier DM with mass $m_{\text{DM}}\gtrsim 250$ GeV will open the new channel $H_2 H_2 \to h h h h $ 
and we can see its small effects in Fig~\ref{dirdelta1-10} and Fig~\ref{dirdelta40-100}.

\begin{figure}
\begin{minipage}{0.39\textwidth}
\includegraphics[width=\textwidth,angle =-90]{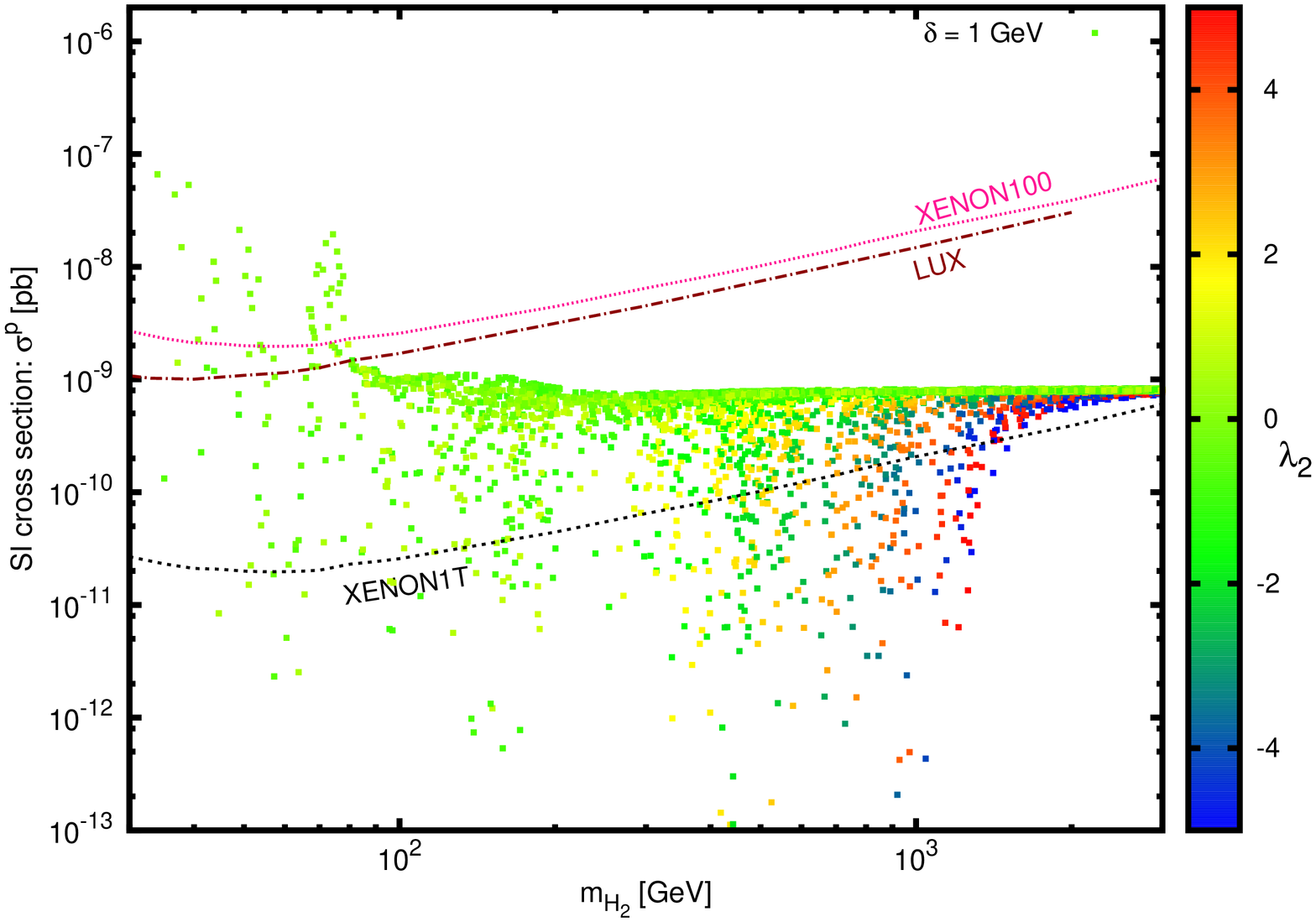}
\end{minipage}
\hspace{2cm}
\begin{minipage}{0.39\textwidth}
\includegraphics[width=\textwidth,angle =-90]{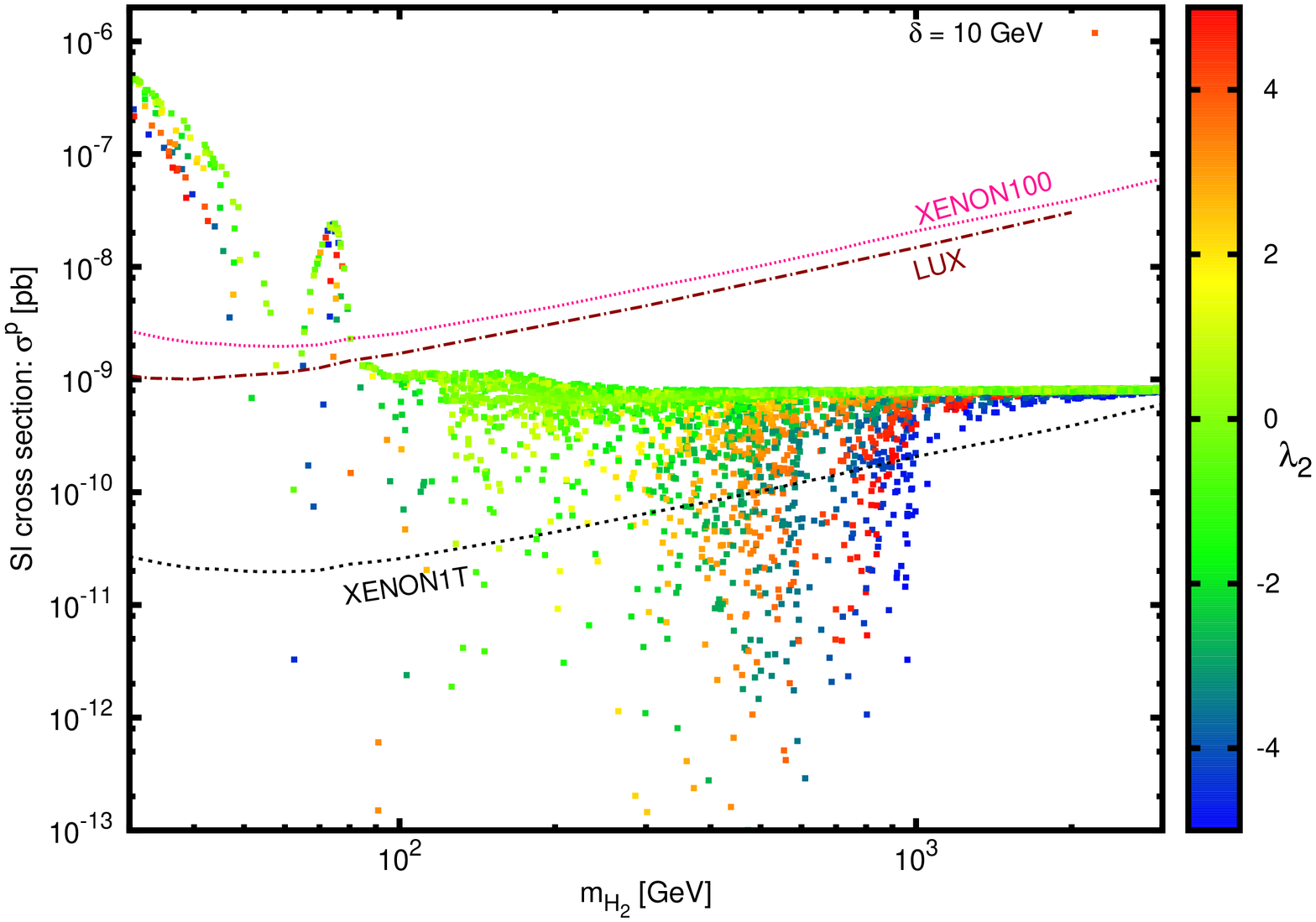}
\end{minipage}
\caption{The same as in Fig.~\ref{dirdelta1-10} 
with $-5 < \lambda_1 < 5$ and $-5 < \lambda_2 < 5$.}
\label{largelanda1}
\end{figure}

\begin{figure}
\begin{minipage}{0.39\textwidth}
\includegraphics[width=\textwidth,angle =-90]{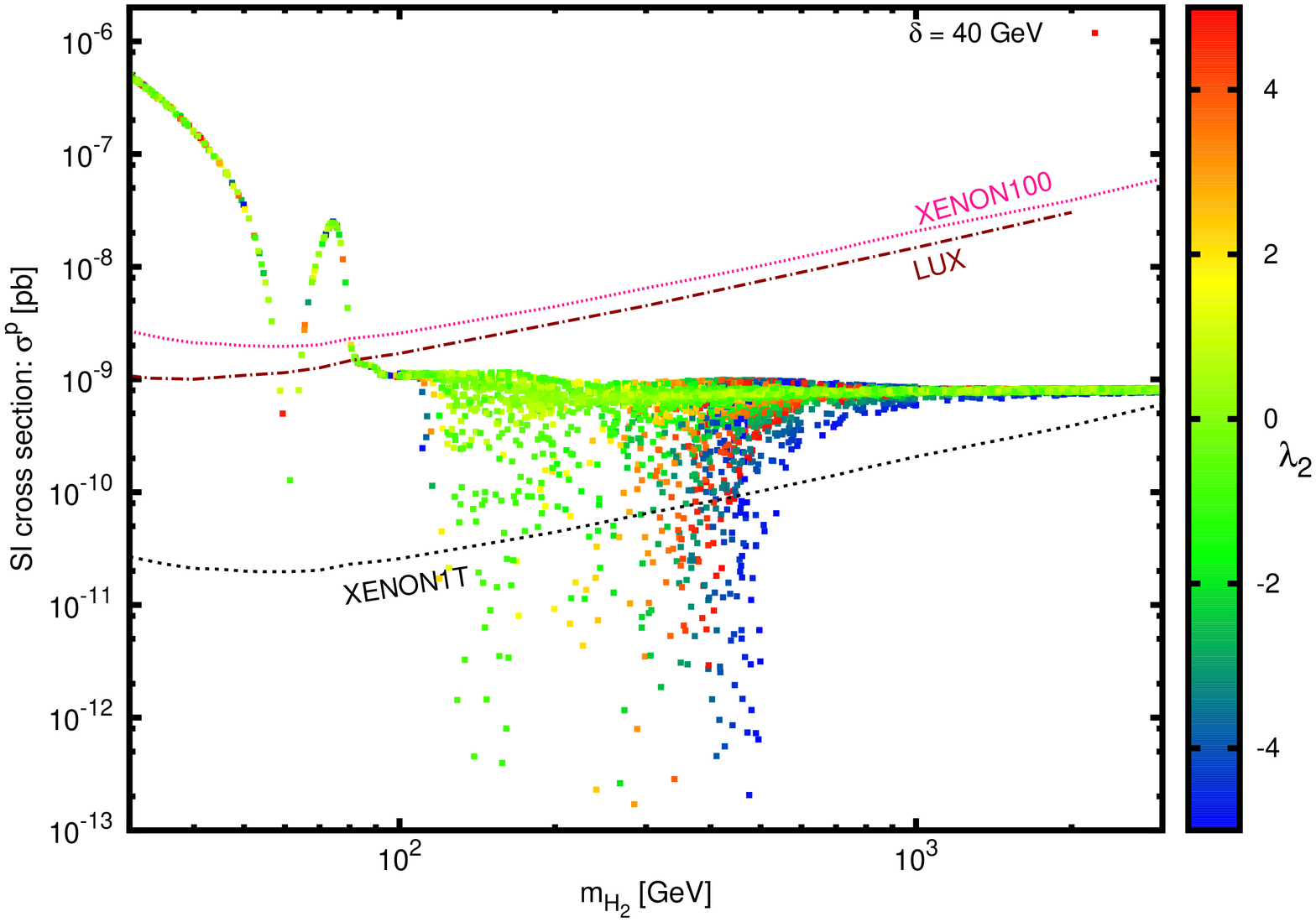}
\end{minipage}
\hspace{2cm}
\begin{minipage}{0.39\textwidth}
\includegraphics[width=\textwidth,angle =-90]{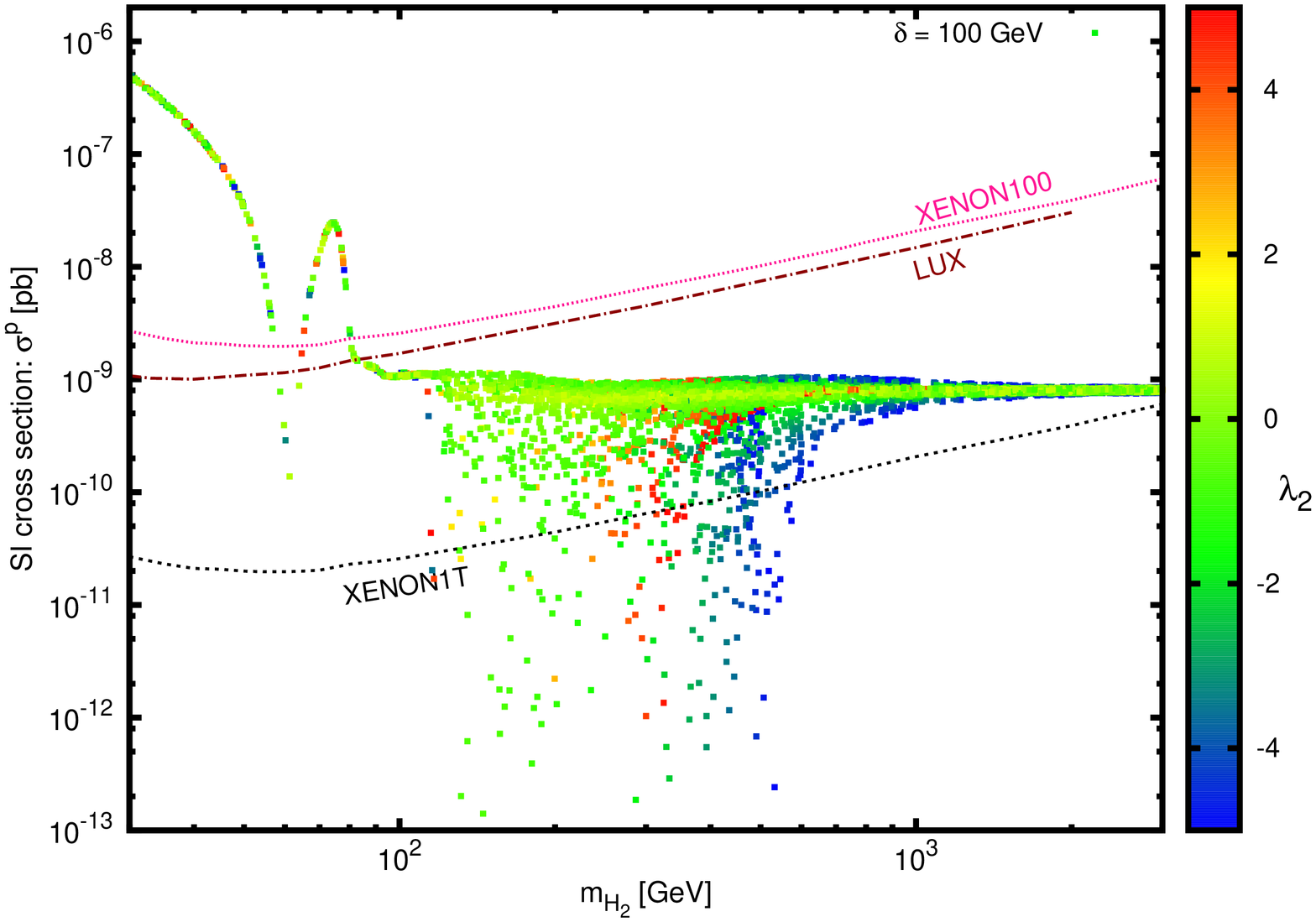}
\end{minipage}
\caption{The same as in Fig.~\ref{dirdelta40-100} with 
$-5 < \lambda_1 < 5$ and $-5 < \lambda_2 < 5$.}
\label{largelanda2}
\end{figure}

We redo our computations with couplings in the 
range $-5 < \lambda_1 < 5$ and $-5 < \lambda_2 < 5$ for $\delta = 1, 10, 40, 100$ GeV. 
Our results given in Fig.~\ref{largelanda1} for $\delta = 1, 10$ GeV indicate that 
for larger values of the couplings $\lambda_1$ and $\lambda_2$, DM candidates 
which can evade XENON1T constraints are extended to masses up to $\sim 1000$ GeV. 
The DM candidates are extended to masses up to $\sim 500$ GeV in case $\delta = 40, 100$ GeV,
as depicted in Fig.~\ref{largelanda2}. 
Comparing our results in Figs.~\ref{largelanda1},\ref{largelanda2} with 
those in Figs.~\ref{dirdelta1-10},\ref{dirdelta40-100}, we realize 
that the effects associated with the 
processes $H_2 H_2 \to h h h$ and $H_2 H_2 \to h h h h $ become sizable 
for $|\lambda_1| \gtrsim 1 $ and $|\lambda_2| \gtrsim 1$.

For the current model, inelastic WIMP-nucleon interaction 
begins contributing for $\delta\sim$ KeV which is far 
smaller from the limits we have considered in this 
work i.e., $\delta\sim$ GeV.

\section{Gamma-ray Emission from DM Annihilation}
\label{gamma-ray-emission}
The analysis of Fermi Large Area Telescope (Fermi-LAT) data \cite{Fermi-LAT1} 
(see \cite{Fermi-LAT2} for the recent report) 
triggered by the authors in \cite{Goodenough:2009gk,Hooper:2010mq} 
and continued by several groups \cite{Agrawal:2014oha,Calore:2014xka,Boyarsky:2010dr,Hooper2011,Abazajian2012,
Hooper:2013rwa,Gordon:2013vta,Abazajian:2014fta,Daylan:2014rsa,Zhou:2014lva} 
revealed an excess in the gamma-ray from the center of the Milky Way 
or Galaxy Center (GC), hence dubbed Galactic Center Excess (GCE). 
The gamma-ray emission produced by the millisecond pulsars 
in the galaxy center can only contribute 
in 5-10\% of the excess
observed \cite{Hooper:2013nhl} \footnote{Although a recent paper \cite{Bartels:2015aea} 
associates GCE to the point-like sources such as millisecond pulsars.}. 
Sources such as cosmic 
ray interactions are disfavored as well \cite{Linden2012}. 

On the other hand, surprisingly the morphology and the 
spectrum of the GCE is well fitted when the dark matter
annihilation into standard model particles is added in 
the background model used in the analyses.
All diffuse background models where include the WIMP 
as a component agree in morphology.
However, the position of the gamma-ray peak in the 
energy spectrum and the mass of the dark matter annihilating 
into SM particles varies by considering the systematic 
uncertainties in the background model \cite{Agrawal:2014oha,Calore:2014xka}.

The dark matter candidate depending on its mass can annihilate 
into leptons, quarks, the gauge bosons and the Higgs boson. The 
gamma-ray is then produced through the cascade decays of these particles to 
neutral pion by the hadronization of the quarks, also through 
the bremsstrahlung of the charged gauge bosons and leptons. 
Among these processes the gamma-ray 
from the pion decay is dominated compared with the gamma 
emission from bremsstrahlung. The differential gamma-ray flux 
produced by a single $W$, $Z$, the Higgs boson and the top quark 
is depicted in Fig.~1 of \cite{Agrawal:2014oha}. 
It can be easily seen that the peak of the spectrum is 
moving towards the higher energies for heavier particles. 

It was believed formerly (see e.g. \cite{Gordon:2013vta,Daylan:2014rsa}) 
that dark matter candidates with masses being only in the 
range of 30 GeV $< m_{\text{DM}} < 50$ GeV  decaying into $\bar{b}b$ give
an acceptable fit with the excess observed in the Fermi data. 
In the recent works 
however it is argued that taking into account 
the systematic uncertainties in the analysis of the Fermi data 
not only the mass range of dark matter for $b\bar{b}$ channel 
is enlarged into 35 GeV $< m_{\text{DM}} < 165$ GeV but also larger dark matter masses 
in annihilation to $WW$, $ZZ$, $hh$, and $t\bar{t}$ 
can be fitted well enough with the data \cite{Agrawal:2014oha}. 
Additionally, it is pointed out in \cite{Calore:2014nla}
that DM mass up to about 74 GeV decaying into $b$ quark pair and also DM annihilation 
into non-relativistic $hh$ can fit well to the Fermi data. 

We show that the gamma-ray excess in our scalar split model 
can be explained well. To this end, we obtain 
the photon flux produced by dark matter annihilation where the allowed values
for the couplings are taken from the viable parameter space.
\begin{figure} 
    \centering
   \includegraphics[scale=.55,angle=0]{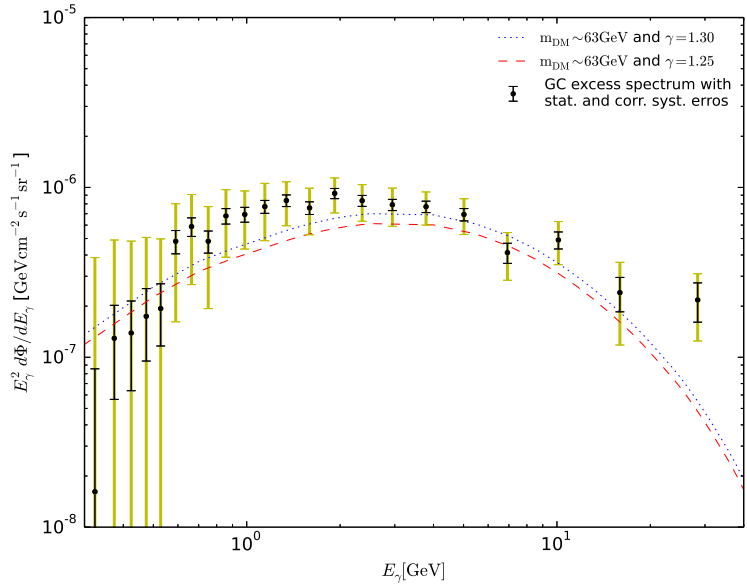}\label{gamma}
    \caption{Shown is the gamma-ray flux multiplied by energy squared 
     from annihilating dark matter computed 
    in the scalar split WIMPs model for dark matter mass $m_{\text{DM}} \sim 63$ 
    GeV and $\delta = 100$ GeV. The black error bars accompanied with correlated 
    systematic errors is the obtained flux from Fermi-LAT data \cite{Calore:2014xka}.}
 \label{gamma}
 \end{figure}

The gamma-ray flux is determined in terms of the annihilation 
cross section $\langle \sigma v \rangle_{\text{ann}}$, the mass 
of the annihilating dark matter $m_{\text{DM}}$, the gamma-ray 
spectrum generated per annihilation $dN_{\gamma}/dE_{\gamma}$ 
and the density of dark matter $\rho$ in the region of 
interest (ROI) is
\begin{equation}
 \frac{d^2\Phi}{dE_{\gamma}d\Omega} = \frac{1}{16\pi}\frac{\langle \sigma v \rangle_{\text{ann}}}{m^{2}_{\text{DM}}}
\frac{dN_{\gamma}}{dE_{\gamma}} \int^{\infty}_{0} dr \rho^{2}(r) \,,
\end{equation} 

The density of dark matter in the Milky Way galaxy is assumed to be spherically 
symmetric. The density distribution is then a
function of $r$ and is described by the generalized Navarro-Frenk-White (NFW) 
halo profile \cite{Navarro:1995iw}:
\begin{equation}
 \rho(r) = \rho_{\odot} (\frac{r_{\odot}}{r})^\gamma \Big(\frac{r_{s}+r_{\odot}}{r_{s}+r} \Big)^{3-\gamma}\,,
\end{equation}
where $r_s=20$ kpc is the scale radius, $\rho_{\odot}=0.3$ GeV/cm$^3$ is the 
local dark matter density at $r_{\odot}=8.5$ kpc and $r$ is the distance 
from the center of the galaxy to the point where the dark matter 
annihilation occurs. The parameter $\gamma$  is the slope 
parameter being $\gamma=1$ for the standard NFW. 
In our calculations we take $\gamma$ within 
the interval $\gamma=1.2-1.3$ used in the literature. 

We find out that within the parameter space confined by relic 
density and direct detection in our DM model,
there can be found regions producing gamma excess that are compatible with the 
fluxes provided by the Fermi data. 
We have used MicrOMEGAs package for computation of the gamma-ray 
flux in our particular model with dark matter mass 
$m_{\text{DM}} \sim 63$ GeV, mass splitting $\delta = 100$ GeV, 
$\lambda_1 = 1.17 \times 10^{-2}$, $\lambda_2 = 6.07 \times 10^{-1}$ and $\sin{\theta} = 0.1$.

In Fig.~\ref{gamma} we present our results for the gamma-ray 
flux multiplied by the gamma energy squared for two slope parameters $\gamma=1.25$ and $\gamma=1.30$.
In this figure it is shown the prediction of the scalar split WIMPs model 
for the gamma excess from annihilating dark matter of mass $\sim$ 63 GeV 
with the total annihilation 
cross section $\langle \sigma  \rangle_{\text{ann}} \sim 4 \times 10^{-26} \text{cm}^{3}\text{s}^{-1}$  
and DM-nucleon elastic scattering cross section $\sigma_{\text{el}} \sim 8 \times 10^{-10}$ pb, 
to be compared with the excess observed from the Fermi-LAT data. 
Comparison made with the data analysis provided by \cite{Calore:2014xka}
at high Galactic latitudes $2^{\circ} \leq |b| \leq 20^{\circ}$ indicates 
the validity of the current model in explaining the Galactic gamma excess.

\section{Conclusions}
\label{sect:conclusions}

In this paper we have employed a simple model of dark matter called 
Scalar Split WIMPs with two scalars $H_1 $ and $H_2$ 
interacting with SM particles through the Higgs portal. 
Depending on the mass splitting $\delta =m_{H_1}-m_{H_2}$ 
and the couplings in the model, 
the decay rate of the heavier scalar $H_1$ changes. 
For the parameter space explored in this work, the $H_1$ decay rate 
is much smaller than the age of the Universe. Therefore we have 
only one scalar $H_2$ that contributes in the DM relic 
abundance. The model possesses seven free parameters out 
of which only five parameters $m_{H{1}}$, $m_{H_{2}}$, $\lambda_{1}$, 
$\lambda_{2}$ and $\theta$ enter into the annihilation and co-annihilation
cross section computations. The mass splitting parameter $\delta$ may 
change the viable parameter space in various computations.

We have examined our model with four observational and experimental 
bounds imposed by invisible Higgs decay, the amount of 
dark matter abundance, the limits put on DM-nucleon cross 
section by direct detection experiments, and the gamma excess found 
by the analyses on the Fermi-LAT data. 

There is an important characteristic for the scalar split WIMP model that distinguishes that from 
the vastly studied singlet scalar models. 
In case we choose $-1 < \lambda_1 < 1 $, $-1 < \lambda_2 < 1 $, 
there can be found viable regions in the parameter space beyond the 
resonant region in the scalar split model with $m_{\text{DM}}$ in the range around $57-200$ GeV for 
$\delta=1$ GeV in Fig. \ref{dirdelta1-10} and around $ 125-200$ GeV for $\delta= 40, 100$ GeV in 
Figs. \ref{dirdelta40-100},
which evade the future experiment bounds on the WIMP-nucleon elastic scattering cross section. 
The viable region is extended to higher DM mass up to $\sim 1000$ GeV 
if we choose $-5 < \lambda_1 < 5 $, $-5 < \lambda_2 < 5$.
For the DM candidates in the viable space that respect the XENON1T and LUX bounds, we 
have inferred that the presence of the co-annihilation (when the mass splitting is small) 
and the mixing effect in the scalar split model play a critical role in 
the new feature so that the model can evade easily the bounds from the future 
direct detection experiments. 

This feature is absent in the singlet scalar model where as 
shown in \cite{Cline:2013gha} the viable region in the parameter space 
is confined in the resonant region with $m_{\text{DM}} $ in the range 
$57-62$ GeV or for $m_{\text{DM}}\gtrsim 4$ TeV by the forthcoming direct detection experiments.

We observe that changing the mass splitting 
$\delta$ has almost no effect on the invisible Higgs decay width. Furthermore, 
the behavior of the relic density for a wide range of dark matter mass has been studied when the mass splitting
takes $\delta=1,4,40,100$ GeV. We observe that the relic density changes considerably 
by varying the mass splitting $\delta$ when $m_{\text{DM}}\lesssim 60 $ GeV
or $m_{\text{DM}}\gtrsim 400$ GeV.

In addition, the scalar split model predicts a gamma-ray excess 
for $m_{\text{DM}}\sim 63$ GeV and $\delta=100$ GeV which is in agreement in morphology 
and spectrum with the excess 
observed out of the Fermi-LAT data. To compute the gamma-ray flux which is produced by 
bremsstrahlung processes and the pion decay created from 
cascade annihilations of dark matter into SM final states, 
we have used the so-called generalized NFW halo profile for the dark matter density
at high Galactic latitudes of the Milky Way galaxy.

\appendix
\section{Annihilation Cross Sections}\label{app}
In this section we present the relevant annihilation cross section formula
for Feynman diagrams with only two particles in the final state.  
The annihilation cross sections into fermion pairs for the dark matter candidate, $H_{2}$ is 
\ba
\label{ann1}
\sigma_{\text{ann}} v_{\text{rel}} ( H_{2} H_{2} \to \bar f f) = 
\frac{N_{c}  m_{f}^2}{\pi} (1-\frac{4m^{2}_{f}}{s})^{\frac{3}{2}}  
\Big[  \frac{(\lambda_{1}\cos^2 \theta+\lambda_{2}\sin^2 \theta - \lambda_{12} \sin \theta \cos \theta)^2}
{(s-m^{2}_{h})^2+m^{2}_{h}\Gamma^{2}_{h}} \Big] \,,
\ea
and for annihilation into gauge bosons $W^{\pm}$ and $Z$ is  

\ba
\label{ann2}
\sigma_{\text{ann}} v_{\text{rel}} ( H_{2} H_{2} \to \bar W^{+} W^{-},ZZ) = 
\frac{1}{2\pi s} 
\Big[  \frac{(\lambda_{1}\cos^2 \theta+\lambda_{2}\sin^2 \theta - \lambda_{12} \sin \theta \cos \theta)^2}
{(s-m^{2}_{h})^2+m^{2}_{h}\Gamma^{2}_{h}} \Big] \times
\nonumber \\&&\hspace{-13.5cm}
 \Big[((s-2m_{W}^2)^2+8m_{W}^2)(1-\frac{4m^{2}_{W}}{s})^{\frac{1}{2}} 
  + \frac{1}{2} ((s-2m_{Z}^2)^2+8m_{W}^2)(1-\frac{4m^{2}_{Z}}{s})^{\frac{1}{2}}  \Big] \,.
\ea
The annihilation cross section for the process $H_2 H_2 \to h h$ involves three Feynman 
diagrams given in Fig.~\ref{darkanni}. The finall result reads
\ba
\label{ann3}
\sigma_{\text{ann}} v_{\text{rel}} ( H_{2} H_{2} \to hh) = \frac{\sqrt{1-4m_h^2/s}}{32\pi^2s} \int d\Omega 
\Big[2 \beta^2 + \frac{72v^4\beta^2 \lambda_{H}^2}{(s-m_{h}^2)^2}+ 
\frac{v^4\alpha^4}{(t-m_{H_1}^2)^2}
\nonumber \\&&\hspace{-13.5cm}
+\frac{v^4\alpha^4}{(u-m_{H_1}^2)^2}+
\frac{16v^4\beta^4}{(t-m_{H_2}^2)^2}
+\frac{16v^4\beta^4}{(u-m_{H_2}^2)^2}+
\frac{16v^2\beta^3}{t-m_{H_2}^2}
+\frac{16v^2\beta^3}{u-m_{H_2}^2}+
\frac{4v^2\beta \alpha^2}{t-m_{H_1}^2}
\nonumber \\&&\hspace{-13.5cm}
+\frac{4v^2\beta \alpha^2}{u-m_{H_1}^2}
-\frac{24v^2\beta^2 \lambda_{H}}{s-m_{h}^2}
-\frac{96v^4\beta^3 \lambda_{H}}{(s-m_{h}^2)(t-m_{H_2}^2)}
-\frac{96v^4\beta^3 \lambda_{H}}{(s-m_{h}^2)(u-m_{H_2}^2)}  
-\frac{24v^4\beta \alpha^2 \lambda_{H}}{(s-m_{h}^2)(t-m_{H_1}^2)}
\nonumber \\&&\hspace{-13.5cm}
-\frac{24v^4\beta \alpha^2 \lambda_{H}}{(s-m_{h}^2)(u-m_{H_1}^2)}  
+\frac{16v^4\beta^4}{(t-m_{H_2}^2)(u-m_{H_2}^2)}
+\frac{v^4\alpha^4}{(t-m_{H_1}^2)(u-m_{H_1}^2)}
+\frac{8v^4 \alpha^2 \beta^2}{(t-m_{H_1}^2)(t-m_{H_2}^2)}
\nonumber \\&&\hspace{-13.5cm}
+\frac{8v^4 \alpha^2 \beta^2}{(t-m_{H_1}^2)(u-m_{H_2}^2)}
+\frac{8v^4 \alpha^2 \beta^2}{(u-m_{H_1}^2)(t-m_{H_2}^2)}
+\frac{8v^4 \alpha^2 \beta^2}{(u-m_{H_1}^2)(u-m_{H_2}^2)}
\Big] \,,
\ea 
where $\alpha = (\lambda_{1}-\lambda_{2}) \sin 2\theta+\lambda_{12} \cos 2\theta$ and  
$\beta = \lambda_{1}\cos^2 \theta+\lambda_{2}\sin^2 \theta - \lambda_{12} \sin \theta \cos \theta$.
In the process $H_2(p_1)~H_2 (p_2)\to h(p_3)~ h(p_4)$, the Mandelstam variables 
are $s = (p_1 +p_2)^2$, $t = (p_1 - p_3)^2$ and $u = (p_1 - p_4)^2$.


\begin{thebibliography}{99}

\bibitem{Ade:2013zuv}
  P.~A.~R.~Ade {\it et al.}  [Planck Collaboration],
  Astron.\ Astrophys.\  {\bf 571} (2014) A16
  [arXiv:1303.5076 [astro-ph.CO]].

\bibitem{Hinshaw:2012aka}
  G.~Hinshaw {\it et al.}  [WMAP Collaboration],
  Astrophys.\ J.\ Suppl.\  {\bf 208} (2013) 19
  [arXiv:1212.5226 [astro-ph.CO]].

\bibitem{Bertone:review}
  G.~Bertone, D.~Hooper and J.~Silk,
  Phys.\ Rept.\  {\bf 405} (2005) 279
  [hep-ph/0404175].

\bibitem{Bergstrom:review}
  L.~Bergstrom,
  Rept.\ Prog.\ Phys.\  {\bf 63} (2000) 793
  [hep-ph/0002126].


\bibitem{Burgess2000}
  C.~P.~Burgess, M.~Pospelov and T.~ter Veldhuis,
  Nucl.\ Phys.\ B {\bf 619} (2001) 709
  [hep-ph/0011335].

\bibitem{McDonald:1993ex}
  J.~McDonald,
  Phys.\ Rev.\ D {\bf 50} (1994) 3637
  [hep-ph/0702143 [HEP-PH]].

\bibitem{Cirelli:2005uq}
  M.~Cirelli, N.~Fornengo and A.~Strumia,
  Nucl.\ Phys.\ B {\bf 753} (2006) 178
  [hep-ph/0512090].


\bibitem{Pospelov:2011-CPfermionic}
  M.~Pospelov and A.~Ritz,
  Phys.\ Rev.\ D {\bf 84} (2011) 113001
  [arXiv:1109.4872 [hep-ph]].

\bibitem{LopezHonorez:2012-CPfermionic}
  L.~Lopez-Honorez, T.~Schwetz and J.~Zupan,
  Phys.\ Lett.\ B {\bf 716} (2012) 179
  [arXiv:1203.2064 [hep-ph]].

\bibitem{Barger:2008jx}
  V.~Barger, P.~Langacker, M.~McCaskey, M.~Ramsey-Musolf and G.~Shaughnessy,
  Phys.\ Rev.\ D {\bf 79} (2009) 015018
  [arXiv:0811.0393 [hep-ph]].


\bibitem{Pospelov:2007mp}
  M.~Pospelov, A.~Ritz and M.~B.~Voloshin,
  Phys.\ Lett.\ B {\bf 662} (2008) 53
  [arXiv:0711.4866 [hep-ph]].


\bibitem{Buckley2014}
  M.~R.~Buckley, D.~Feld and D.~Goncalves,
  arXiv:1410.6497 [hep-ph].

\bibitem{Akerib:LUX}
  D.~S.~Akerib {\it et al.}  [LUX Collaboration],
  Phys.\ Rev.\ Lett.\  {\bf 112} (2014) 091303
  [arXiv:1310.8214 [astro-ph.CO]].

\bibitem{Aprile:XENON100}
  E.~Aprile {\it et al.}  [XENON100 Collaboration],
  Phys.\ Rev.\ Lett.\  {\bf 109} (2012) 181301
  [arXiv:1207.5988 [astro-ph.CO]].


\bibitem{Beltran:2010ww}
  M.~Beltran, D.~Hooper, E.~W.~Kolb, Z.~A.~C.~Krusberg and T.~M.~P.~Tait,
  JHEP {\bf 1009} (2010) 037
  [arXiv:1002.4137 [hep-ph]].

\bibitem{Goodman:2010ku}
  J.~Goodman, M.~Ibe, A.~Rajaraman, W.~Shepherd, T.~M.~P.~Tait and H.~B.~Yu,
  Phys.\ Rev.\ D {\bf 82} (2010) 116010
  [arXiv:1008.1783 [hep-ph]].

\bibitem{Bai:2010hh}
  Y.~Bai, P.~J.~Fox and R.~Harnik,
  JHEP {\bf 1012} (2010) 048
  [arXiv:1005.3797 [hep-ph]].

\bibitem{Belanger:invisible}
  G.~Belanger, B.~Dumont, U.~Ellwanger, J.~F.~Gunion and S.~Kraml,
  Phys.\ Lett.\ B {\bf 723} (2013) 340
  [arXiv:1302.5694 [hep-ph]].


\bibitem{ArkaniHamed:2002pa}
  N.~Arkani-Hamed, A.~G.~Cohen, T.~Gregoire and J.~G.~Wacker,
  JHEP {\bf 0208} (2002) 020
  [hep-ph/0202089].
  
\bibitem{Barger:2007im}
  V.~Barger, P.~Langacker, M.~McCaskey, M.~J.~Ramsey-Musolf and G.~Shaughnessy,
  Phys.\ Rev.\ D {\bf 77} (2008) 035005
  [arXiv:0706.4311 [hep-ph]].
 
\bibitem{Cline:2013gha}
  J.~M.~Cline, K.~Kainulainen, P.~Scott and C.~Weniger,
  Phys.\ Rev.\ D {\bf 88} (2013) 055025
  [arXiv:1306.4710 [hep-ph]].
 
 
 
 \bibitem{Boehm:2014hva}
  C.~Boehm, M.~J.~Dolan, C.~McCabe, M.~Spannowsky and C.~J.~Wallace,
  JCAP {\bf 1405} (2014) 009
  [arXiv:1401.6458 [hep-ph]].

\bibitem{Berlin:2014tja}
  A.~Berlin, D.~Hooper and S.~D.~McDermott,
  Phys.\ Rev.\ D {\bf 89} (2014) 115022
  [arXiv:1404.0022 [hep-ph]].

\bibitem{Boehm:2014bia}
  C.~Boehm, M.~J.~Dolan and C.~McCabe,
  Phys.\ Rev.\ D {\bf 90} (2014) 023531
  [arXiv:1404.4977 [hep-ph]].

\bibitem{Ko:2014gha}
  P.~Ko, W.~I.~Park and Y.~Tang,
  JCAP {\bf 1409} (2014) 013
  [arXiv:1404.5257 [hep-ph]].

\bibitem{Abdullah:2014lla}
  M.~Abdullah, A.~DiFranzo, A.~Rajaraman, T.~M.~P.~Tait, P.~Tanedo and A.~M.~Wijangco,
  Phys.\ Rev.\ D {\bf 90} (2014) 3,  035004
  [arXiv:1404.6528 [hep-ph]].

\bibitem{Berlin:2014pya}
  A.~Berlin, P.~Gratia, D.~Hooper and S.~D.~McDermott,
  Phys.\ Rev.\ D {\bf 90} (2014) 015032
  [arXiv:1405.5204 [hep-ph]].

\bibitem{Cline:2014dwa}
  J.~M.~Cline, G.~Dupuis, Z.~Liu and W.~Xue,
  JHEP {\bf 1408} (2014) 131
  [arXiv:1405.7691 [hep-ph]].

\bibitem{Wang:2014elb}
  L.~Wang,
  arXiv:1406.3598 [hep-ph].

\bibitem{Cheung:2014lqa}
  C.~Cheung, M.~Papucci, D.~Sanford, N.~R.~Shah and K.~M.~Zurek,
  Phys.\ Rev.\ D {\bf 90} (2014) 075011
  [arXiv:1406.6372 [hep-ph]].

\bibitem{Balazs:2014jla}
  C.~Balázs and T.~Li,
  Phys.\ Rev.\ D {\bf 90} (2014) 055026
  [arXiv:1407.0174 [hep-ph]].

\bibitem{Huang:2014cla}
  J.~Huang, T.~Liu, L.~T.~Wang and F.~Yu,
  Phys.\ Rev.\ D {\bf 90} (2014) 115006
  [arXiv:1407.0038 [hep-ph]].

\bibitem{Ghorbani:2014qpa}
  K.~Ghorbani,
  JCAP {\bf 1501} (2015) 015
  [arXiv:1408.4929 [hep-ph]].


\bibitem{Banik:2014eda}
  A.~D.~Banik and D.~Majumdar,
  arXiv:1408.5795 [hep-ph].

\bibitem{Borah:2014ska}
  D.~Borah and A.~Dasgupta,
  arXiv:1409.1406 [hep-ph].

\bibitem{Cahill-Rowley:2014ora}
  M.~Cahill-Rowley, J.~Gainer, J.~Hewett and T.~Rizzo,
  arXiv:1409.1573 [hep-ph].


\bibitem{Guo:2014gra}
  J.~Guo, J.~Li, T.~Li and A.~G.~Williams,
  arXiv:1409.7864 [hep-ph].

\bibitem{Dolan:2014ska}
  M.~J.~Dolan, C.~McCabe, F.~Kahlhoefer and K.~Schmidt-Hoberg,
  arXiv:1412.5174 [hep-ph].

\bibitem{Biswas:2014hoa}
  A.~Biswas,
  arXiv:1412.1663 [hep-ph].

  \bibitem{Modak:2013jya}
  K.~P.~Modak, D.~Majumdar and S.~Rakshit,
  JCAP {\bf 1503} (2015) 011
  [arXiv:1312.7488 [hep-ph]].
  
  
\bibitem{Cao:2014efa}
  J.~Cao, L.~Shang, P.~Wu, J.~M.~Yang and Y.~Zhang,
  arXiv:1410.3239 [hep-ph].

\bibitem{Bell:2014xta}
  N.~F.~Bell, S.~Horiuchi and I.~M.~Shoemaker,
  Phys.\ Rev.\ D {\bf 91} (2015) 023505
  [arXiv:1408.5142 [hep-ph]].

\bibitem{Detmold:2014qqa}
  W.~Detmold, M.~McCullough and A.~Pochinsky,
  Phys.\ Rev.\ D {\bf 90} (2014) 11,  115013
  [arXiv:1406.2276 [hep-ph]].

\bibitem{Cheung:2014tha}
  K.~Cheung, W.~C.~Huang and Y.~L.~S.~Tsai,
  arXiv:1411.2619 [hep-ph].

\bibitem{Ko:2014loa}
  P.~Ko and Y.~Tang,
  JCAP {\bf 1501} (2015) 023
  [arXiv:1407.5492 [hep-ph]].

\bibitem{Alvares:2012qv}
  J.~D.~Ruiz-Alvarez, C.~A.~de S.Pires, F.~S.~Queiroz, D.~Restrepo and P.~S.~Rodrigues da Silva,
  Phys.\ Rev.\ D {\bf 86} (2012) 075011
  [arXiv:1206.5779 [hep-ph]].

\bibitem{Basak:2014sza}
  T.~Basak and T.~Mondal,
  arXiv:1405.4877 [hep-ph].

\bibitem{Martin:2014sxa}
  A.~Martin, J.~Shelton and J.~Unwin,
  Phys.\ Rev.\ D {\bf 90} (2014) 10,  103513
  [arXiv:1405.0272 [hep-ph]].

\bibitem{Hardy:2014dea}
  E.~Hardy, R.~Lasenby and J.~Unwin,
  JHEP {\bf 1407} (2014) 049
  [arXiv:1402.4500 [hep-ph]].



\bibitem{Marshall:2011mm}
  G.~Marshall and R.~Primulando,
  JHEP {\bf 1105} (2011) 026
  [arXiv:1102.0492 [hep-ph]].

\bibitem{Okada:2013bna}
  N.~Okada and O.~Seto,
  Phys.\ Rev.\ D {\bf 89} (2014) 043525
  [arXiv:1310.5991 [hep-ph]].


\bibitem{Liu:2014cma}
  J.~Liu, N.~Weiner and W.~Xue,
  arXiv:1412.1485 [hep-ph].


\bibitem{Freytsis:2014sua}
  M.~Freytsis, D.~J.~Robinson and Y.~Tsai,
  arXiv:1410.3818 [hep-ph].

\bibitem{Lacroix:2014eea}
  T.~Lacroix, C.~Boehm and J.~Silk,
  Phys.\ Rev.\ D {\bf 90} (2014) 4,  043508
  [arXiv:1403.1987 [astro-ph.HE]].


\bibitem{Agrawal:2014oha}
  P.~Agrawal, B.~Batell, P.~J.~Fox and R.~Harnik,
  arXiv:1411.2592 [hep-ph].

\bibitem{Calore:2014xka}
  F.~Calore, I.~Cholis and C.~Weniger,
  arXiv:1409.0042 [astro-ph.CO].

\bibitem{Calore:2014nla}
  F.~Calore, I.~Cholis, C.~McCabe and C.~Weniger,
  arXiv:1411.4647 [hep-ph].

\bibitem{Griest:1990kh}
  K.~Griest and D.~Seckel,
  Phys.\ Rev.\ D {\bf 43} (1991) 3191.
  doi:10.1103/PhysRevD.43.3191


\bibitem{Edsjo:1997bg}
  J.~Edsjo and P.~Gondolo,
  Phys.\ Rev.\ D {\bf 56} (1997) 1879
  doi:10.1103/PhysRevD.56.1879
  [hep-ph/9704361].

\bibitem{Belanger:2004yn}
  G.~Belanger, F.~Boudjema, A.~Pukhov and A.~Semenov,
  Comput.\ Phys.\ Commun.\  {\bf 174} (2006) 577
  doi:10.1016/j.cpc.2005.12.005
  [hep-ph/0405253].



\bibitem{Belyaev:CalcHEP}
  A.~Belyaev, N.~D.~Christensen and A.~Pukhov,
  Comput.\ Phys.\ Commun.\  {\bf 184} (2013) 1729
  [arXiv:1207.6082 [hep-ph]].

\bibitem{Semenov:LanHEP}
  A.~Semenov,
  arXiv:1005.1909 [hep-ph].

\bibitem{Belanger:MICRO}
  G.~Belanger, F.~Boudjema, A.~Pukhov and A.~Semenov,
  Comput.\ Phys.\ Commun.\  {\bf 185} (2014) 960
  [arXiv:1305.0237 [hep-ph]].




\bibitem{Heinemeyer:2013}
  S.~Heinemeyer {\it et al.}  [LHC Higgs Cross Section Working Group Collaboration],
  arXiv:1307.1347 [hep-ph].
 
  
  
  
\bibitem{Ellis:2008}
  J.~R.~Ellis, K.~A.~Olive and C.~Savage,
  Phys.\ Rev.\ D {\bf 77} (2008) 065026
  [arXiv:0801.3656 [hep-ph]].

  
\bibitem{Belanger:2008-Direct}
  G.~Belanger, F.~Boudjema, A.~Pukhov and A.~Semenov,
  Comput.\ Phys.\ Commun.\  {\bf 180} (2009) 747
  [arXiv:0803.2360 [hep-ph]].
  
    
\bibitem{Nihei:2004}
  T.~Nihei and M.~Sasagawa,
  Phys.\ Rev.\ D {\bf 70} (2004) 055011
   [Erratum-ibid.\ D {\bf 70} (2004) 079901]
  [hep-ph/0404100].


\bibitem{Ellis:2000}
  J.~R.~Ellis, A.~Ferstl and K.~A.~Olive,
  Phys.\ Lett.\ B {\bf 481} (2000) 304
  [hep-ph/0001005].

\bibitem{Crivellin:2013}
  A.~Crivellin, M.~Hoferichter and M.~Procura,
  Phys.\ Rev.\ D {\bf 89} (2014) 054021
  [arXiv:1312.4951 [hep-ph]].



\bibitem{Fermi-LAT1}
  R.~Rando [Fermi LAT Collaboration],
  arXiv:0907.0626 [astro-ph.IM].

\bibitem{Fermi-LAT2}
Fermi-LAT Collaboration, S. Murgia, ``Observation of the High Energy Gamma-ray Emission Towards 
the Galactic Center''.
http://fermi.gsfc.nasa.gov/science/mtgs/symposia/2014/program/08-Murgia.pdf,
(October 2014).

\bibitem{Goodenough:2009gk}
  L.~Goodenough and D.~Hooper,
  arXiv:0910.2998 [hep-ph].
  
 
\bibitem{Hooper:2010mq}
  D.~Hooper and L.~Goodenough,
  Phys.\ Lett.\ B {\bf 697} (2011) 412
  [arXiv:1010.2752 [hep-ph]].
  
  
\bibitem{Boyarsky:2010dr}
  A.~Boyarsky, D.~Malyshev and O.~Ruchayskiy,
  Phys.\ Lett.\ B {\bf 705} (2011) 165
  [arXiv:1012.5839 [hep-ph]].
\bibitem{Hooper2011} 
  D.~Hooper and T.~Linden,
  Phys.\ Rev.\ D {\bf 84} (2011) 123005
  [arXiv:1110.0006 [astro-ph.HE]].


\bibitem{Abazajian2012}
  K.~N.~Abazajian and M.~Kaplinghat,
  Phys.\ Rev.\ D {\bf 86} (2012) 083511
  [arXiv:1207.6047 [astro-ph.HE]].


\bibitem{Hooper:2013rwa}
  D.~Hooper and T.~R.~Slatyer,
  Phys.\ Dark Univ.\  {\bf 2} (2013) 118
  [arXiv:1302.6589 [astro-ph.HE]].
  
\bibitem{Gordon:2013vta}
  C.~Gordon and O.~Macias,
  Phys.\ Rev.\ D {\bf 88} (2013) 083521
   [Erratum-ibid.\ D {\bf 89} (2014) 4,  049901]
  [arXiv:1306.5725 [astro-ph.HE]].
  
\bibitem{Abazajian:2014fta}
  K.~N.~Abazajian, N.~Canac, S.~Horiuchi and M.~Kaplinghat,
  Phys.\ Rev.\ D {\bf 90} (2014) 023526
  [arXiv:1402.4090 [astro-ph.HE]].
  
\bibitem{Daylan:2014rsa}
  T.~Daylan, D.~P.~Finkbeiner, D.~Hooper, T.~Linden, S.~K.~N.~Portillo, N.~L.~Rodd and T.~R.~Slatyer,
  arXiv:1402.6703 [astro-ph.HE].
  
\bibitem{Zhou:2014lva}
  B.~Zhou, Y.~F.~Liang, X.~Huang, X.~Li, Y.~Z.~Fan, L.~Feng and J.~Chang,
  arXiv:1406.6948 [astro-ph.HE].

\bibitem{Hooper:2013nhl}
  D.~Hooper, I.~Cholis, T.~Linden, J.~Siegal-Gaskins and T.~Slatyer,
  Phys.\ Rev.\ D {\bf 88} (2013) 083009
  [arXiv:1305.0830 [astro-ph.HE]].

\bibitem{Linden2012}
  T.~Linden, E.~Lovegrove and S.~Profumo,
  Astrophys.\ J.\  {\bf 753} (2012) 41
  [arXiv:1203.3539 [astro-ph.HE]].

\bibitem{Navarro:1995iw}
  J.~F.~Navarro, C.~S.~Frenk and S.~D.~M.~White,
  Astrophys.\ J.\  {\bf 462} (1996) 563
  [astro-ph/9508025].
  

  
  
\bibitem{Bartels:2015aea}
  R.~Bartels, S.~Krishnamurthy and C.~Weniger,
  arXiv:1506.05104 [astro-ph.HE].



\end{thebibliography}

\end{document}